\documentclass{article}

\usepackage{arxiv}
\usepackage{tikz}
\usepackage[utf8]{inputenc} 
\usepackage[T1]{fontenc}    
\usepackage{hyperref}       
\usepackage{url}            
\usepackage{booktabs}       
\usepackage{paralist}
\usepackage{amssymb}
\usepackage{amsfonts}       
\usepackage{nicefrac}       
\usepackage{microtype}      
\usepackage{lipsum}		
\usepackage{graphicx}
\usepackage{natbib}
\usepackage{doi,xcolor}
\graphicspath{{./images}}
\usepackage{caption,subcaption}
\usepackage{amsmath,bm}
\usepackage{appendix}

\newcommand{\p}{\mathbf{p}}  

\newcommand{\dz}{d^z} 
\newcommand{\Dz}{D^z}
\newcommand{\dr}{d^r} 
\newcommand{\Dr}{D^r}
\newcommand{\gz}{g^z} 
\newcommand{\Gz}{G^z}
\newcommand{\gr}{g^r} 
\newcommand{\Gr}{G^r}
\newcommand{\Grz}{G^{r,z}}
\newcommand{\grz}{g^{r,z}}
\newcommand{\nz}{\epsilon^z} 
\newcommand{\Nz}{\varepsilon^z}
\newcommand{\nr}{\epsilon^r} 
\newcommand{\Nr}{\varepsilon^r}
\newcommand{\Nt}{\varepsilon}
\newcommand{\nt}{\epsilon}
\newcommand{\cc}[1]{\overline{#1}}
\newcommand{\fastbz}{\psi} 
\newcommand{\delt}{\delta t} 

\newcommand{\rec}{\mathbf{r}} 
\newcommand{\vir}{\mathbf{v}} 

\newcommand{\qv}{\bm{\epsilon}}  

\newcommand{\n}{n} 
\newcommand{\Nn}{N} 
\newcommand{\ncc}{\gamma}
\newcommand{\drec}{\mathbf{r}^{\prime}_j} 
\newcommand{\rrec}{\mathbf{r}^{\text{true}}_j} %

\title{Virtual Receiver Functions via Conditional Diffusion Transformers for Robust Crustal Imaging}

\author{ \href{https://orcid.org/0009-0007-2912-8242}{\includegraphics[scale=0.06]{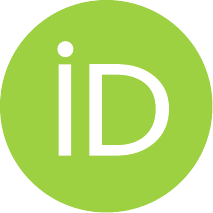}\hspace{1mm}Tiente R. Koireng} \\
	Centre for Earth Sciences\\
	Indian Institute of Science\\
	Bengaluru, India 560012 \\
	\texttt{tienterk@iisc.ac.in} \\
	\And
     \href{https://orcid.org/0000-0003-4081-8969}{\includegraphics[scale=0.06]{orcid.pdf}\hspace{1mm}Priyanshu Gupta} \\
	Centre for Earth Sciences\\
	Indian Institute of Science\\
	Bengaluru, India 560012 \\
	\texttt{gpriyanshu@iisc.ac.in} \\
	   \AND
       \href{https://orcid.org/0000-0003-4081-8969}{\includegraphics[scale=0.06]{orcid.pdf}\hspace{1mm}Pawan Bharadwaj} \\
	Centre for Earth Sciences\\
	Indian Institute of Science\\
	Bengaluru, India 560012 \\
	\texttt{pawan@iisc.ac.in} \\
}

\renewcommand{\shorttitle}{Enhanced Receiver Function Imaging of Crustal Structures}

\hypersetup{
pdftitle={Enhanced Receiver Function Imaging Of Crustal Structures},
pdfauthor={T~Rengneichuong~Koireng, Pawan~Bharadwaj},
pdfkeywords={Receiver functions, Diffusion transformers, Conditional generation, Deconvolution, Crustal imaging, Crustal anisotropy, Subduction},
}

\begin{document}

\maketitle

\begin{abstract}
\label{sec:abstract}

Receiver functions (RFs) are widely used to image crustal and upper-mantle structure, and their variation with backazimuth and epicentral distance contains key information about layering and azimuthal anisotropy.
In practice, however, RFs are contaminated by nuisance effects due to unknown earthquake source signatures and seismic noise, which obstruct the reliable extraction of crustal structure information. In addition, sparse availability of RFs across backazimuths and epicentral distances leads to biased anisotropy estimates.
To address these issues, we use diffusion models to learn the distribution of RFs conditioned on backazimuth, epicentral distance, and station location. The key insight that allows us to suppress nuisance effects is that RFs from earthquakes with similar backazimuths and epicentral distances share coherent crustal effects while exhibiting variable nuisance effects.
We show that diffusion models can generate high-quality virtual radial and transverse RFs with reduced nuisance effects, which can be used to reliably extract crustal structure information. Our models interpolate RFs over both backazimuth and epicentral distance, allowing us to create virtual RFs without gaps in these conditions, which in turn facilitates interpretation of crustal anisotropy and layering geometry.
%
%
We validate the method on synthetic RFs with realistic non-Gaussian noise, where virtual RFs achieve higher correlation with true RFs than traditional linear or phase-weighted averaging. When applied to the Cascadia Subduction Zone, virtual radial RFs clearly image all scattered S-waves from the dipping subducting slab, with improved phase clarity and backazimuthal coverage compared to previous studies. In southern California, anisotropy parameters derived from virtual RFs are spatially coherent and geologically plausible, given the geometry of the regional faults.
In summary, unlike traditional methods that discard low signal-to-noise ratio RFs, our approach utilizes all earthquakes regardless of quality, enabling full backazimuth and epicentral distance coverage essential for robust, automated RF analysis of temporary and permanent seismic networks.
\end{abstract}


\section{Introduction}
\label{sec:intro}

Receiver functions (RFs), computed using several teleseismic earthquakes, provide a powerful method for investigating crustal and upper mantle structures below a particular receiver. 
This study focuses on analyzing the changes in the RF relative to the backazimuth and the distance to the teleseismic earthquake, which are essential for accurate subsurface characterization. Careful analysis of these variations and comparison with forward modeling results~\citep{levin2008seismic, Savage2007, liu2015crustal} can reveal important information about the orientation and magnitude of Earth's anisotropy, as well as the 3-D geometry of subsurface features, providing researchers with valuable insights into the tectonic evolution and dynamics of a given region.
To illustrate this emphasis, we consider the systematic variation of the PS phase (forward-scattered S waves from the Moho caused by an incident P wave) with respect to backazimuth and distance.
Numerous studies have extracted the fast-polarization direction and magnitude of crustal anisotropy from the arrival of PS within RFs~\citep{McNamara1993, Bianchi2010, Zheng2018}. 
Similarly, 
backazimuth variations of the PS arrival in both radial and transverse RFs have been employed to calculate the dip azimuth, dip angle, and depth of subducting slabs~\citep{ShiomiK}. 
~\cite{zhu2000moho} employed RFs generated by earthquakes that occur at various epicentral distances to determine the thickness of the crust and the P to S velocity ratio through semblance analysis.


To analyze RF variations with respect to backazimuth and epicentral distance, one must be able to compare RFs from multiple earthquakes. However, this task can be challenging due to the presence of pseudo-random \emph{nuisance} effects.
The nuisance effects are unique to each earthquake and it should be emphasized that the common assumption that these effects can be modeled as random Gaussian noise frequently proves inadequate~\citep{kolb2014receiver,Bodin2014Inversion,bona1998variance}.
To be precise, 
nuisance effects are unwanted signals in RFs that arise through two distinct pathways during the deconvolution of the radial with the vertical component of the seismogram. First, as deconvolution is ill-posed, it is susceptible to contamination by background seismic noise (owing to natural and human
activities). 
This means that insufficient regularization in deconvolution amplifies the seismic noise linked to the spectral zeros of the source signature, which causes nuisance effects~\citep{Akuhara2019}. Conversely, when regularizing the deconvolution, nuisance effects appear as biases that are specific to the regularization method.
Secondly, deconvolution, which is associated with cross-correlation, results in crosstalk between the scattered waves from crustal structures and seismic noise in the vertical and radial components.
Crosstalk increases interference and makes the interpretation of converted seismic phases related to crustal structures (referred to as crustal effects) more challenging.
%
%

Various researchers~\citep{Park2000,Zhang2022,zhang2024crustal} have recognized that pseudo-random nuisance effects make it difficult to interpret receiver functions, highlighting the need to address these effects to achieve the accurate extraction of crustal effects.
A strategy to improve the quality of RFs involves careful selection and identification of usable earthquakes through automated processes~\citep{Crotwell2005,Yang2016} and machine learning frameworks~\cite{Gong2022,Krueger2021,sabermahani2024deepRFQC}. However, removing earthquakes will decrease azimuthal coverage. 
Therefore, most of the methods focus on reducing nuisance effects\footnote{The nuisance effects cannot be effectively removed through standard bandpass filtering techniques, as they share a frequency band similar to that of crustal converted phases.} in RFs. 
In the literature, these methods can be classified into two categories.

The first category corresponds to simple averaging methods, where multiple RFs are averaged to improve the signal-to-noise ratio (SNR). Examples include
\begin{enumerate}
    \item averaging multiple earthquakes with a similar backazimuth and epicentral distance~\citep{Gurrola1995,Levin1997,Hu2015,Bloch2023};
    \item weighted averaging techniques, such as the frequency-domain method proposed by ~\cite{Park2000}, which can assess the level of nuisance effects in a RF.
    \item phase-weighted averaging~\citep{Schimmel1997} of multiple RFs with similar backazimuth and epicentral distance.
\end{enumerate}
The methods in this category
are efficient when
dealing with 1) permanent stations that record numerous earthquakes;
2) a uniform earthquake distribution across backazimuth and epicentral distance.
%
%
In practice, the earthquake distribution is non-uniform, meaning that some azimuthal or epicentral-distance bins have more earthquakes than others, which can lead to less efficient averaging for certain bins~\citep{Levin1997,Ozakin2015,Bloch2023}.
The selection of weights in
weighted averaging methods can be subjective.
In this work, we average multiple network-generated virtual RFs generated for a specific condition defined by backazimuth, epicentral distance and station location to reduced nuisance effects.



Model-based techniques comprise the second category of methods that aim to minimize nuisance effects. To improve RF quality, these methods rely on assumptions about the Earth's velocity structure.
They apply constraints to a gather of available RFs, ordered according to either backazimuth or epicentral distance. 
Examples include
\begin{enumerate}
    \item methods using sparsity-promoting Randon transforms~\citep{Olugboji2023}, curvelet transforms ~\citep{chen2019denoising}, and singular spectrum analysis~\citep{Dokht2016} assume a laterally homogeneous Earth model to filter linear events from the RF gather;
    \item supervised deep learning techniques~\citep{Wang2022} are trained on crustal velocity models drawn from the prior distribution to output crustal thickness and the P-to-S wave velocity ratio from the RFs; 
    \item common conversion point (CCP) averaging~\citep{Zhu2000} of the RF gather, 
    commonly used for imaging the Moho and mantle discontinuities, involves time-depth conversion based on a large-scale velocity model.
\end{enumerate}
Methods in this category are unsuccessful when applied to intricate geological environments such as subduction zones, which are different from the assumed (simpler) models. These methods also require dense spatial coverage of earthquake events, which is challenging for temporary seismic stations.
CCP averaging may result in artifacts~\citep{Zheng2014} due to mismatches between the assumed velocity model and the Earth's true model.
%
An important drawback of supervised deep learning is the difficulty in assessing its
performance on real seismic data, which originates from a distribution different from that assumed during training.
Our unsupervised method does not require Earth's velocity model and is trained directly on real RFs, making it more robust to complex geological settings and applicable to temporary seismic stations with limited earthquake coverage.

This paper aims to train neural networks on RFs from various earthquakes and then \emph{synthesize} new virtual RFs with minimal nuisance effects across all backazimuths and epicentral distances.
We demonstrate that these virtual RFs retain crustal effects, enabling a reliable interpretation of RFs in complex geology and an accurate estimation of crustal anisotropy parameters.
In computer vision, 
denoising diffusion models have been widely used to generate new images by learning the distribution of the training images~\citep{sohl2015deep,ho2020denoising,dhariwal2021diffusion}. Images are progressively corrupted by adding noise in many small steps until they become  white noise. A neural network is then trained to learn the reverse denoising process at each step, thereby modeling the probability distribution of the training images.  
Diffusion models show strong potential for denoising seismic waveforms, as demonstrated by their superior performance compared with earlier generative approaches such as GANs and VAEs. Recent work indicates that diffusion models also surpass previous deep learning–based methods for seismic denoising~\citep{Durall2023DeepDiffusionSeismic,Trappolini2024ColdDiffusion}.
In addition, diffusion models are increasingly being applied beyond denoising, including for earthquake waveform generation and microseismic source  location estimation ~\citep{jung2025broadband,palgunadi2025highresolutionseismicwaveform,wamriew2025diffusion}.
Diffusion models with transformers (DiT)~\citep{peebles2023scalablediffusionmodelstransformers} have proven highly effective for text-to-image generation, where text prompts serve as conditioning parameters that steer the model to produce images with specific content or characteristics. 

In this study, we train a diffusion transformer on RFs, using backazimuth, epicentral distance, and station location as conditioning parameters. The diffusion transformer is trained to approximate the conditional probability distribution of the nuisance effects and crustal effects given these condition parameters and generate realistic, condition-specific RFs. After training, we generate multiple RFs for a specific condition defined by backazimuth, epicentral distance and station location, and then average these realizations to obtain a virtual RF with reduced nuisance effects.
To validate our methodology, we applied it to RFs derived from synthetic seismograms with real earthquake source signatures and ambient seismic noise. We considered a crustal model with a dipping anisotropic layer. RFs for a certain range of backazimuth and epicentral distance were omitted from the training data and then virtual RFs were generated for these conditions. 
 The quality of the virtual RFs was evaluated by computing the normalized correlation coefficient relative to the true RFs. In general, the virtual RFs showed higher correlation than RFs obtained through linear and phase-weighted stacking. Virtual RFs generated within backazimuth gaps also correlated well with the true RFs, indicating that our model can effectively interpolate along backazimuth.

We tested the performance of our approach in complex geological settings, focusing on the Cascadia subduction zone. 
The Cascadia Subduction Zone, where the Juan de Fuca oceanic plate is subducted beneath the North American continental plate, has drawn significant research attention, similar to other subduction zones. This is due to its complex slab morphology, association with seismic hazards, and geodynamic processes~\citep{Langston1979,Nicholson2005,Bloch2023}.
RF studies have been widely utilized to characterize crustal structures in subduction zones~\citep{Cassidy1995,Bloch2023,Savage2007,ShiomiK}.
However, the complex structural settings in the subduction zones, which include low-velocity zones, partial melting, and accretionary prisms, present significant obstacles in the extraction of crustal and upper mantle structures from RFs.
Virtual radial RFs from Cascadia clearly image all the scattered S-waves associated with velocity contrasts in the subducting oceanic slab, consistent with the observations of~\cite{Bloch2023}.

Southern California provides an ideal complementary test: rather than imaging discrete velocity discontinuities, we seek to extract azimuthal anisotropy of the crust.
 Multiple studies have been conducted to investigate the crustal azimuthal anisotropy and its relationship with the seismogenic processes.
Anisotropy measurements using RFs have revealed a complex pattern of fast-axis orientations around the San Andreas Fault (SAF)~\citep{Porter2011,Audet2015}. Using P waves from local earthquakes, \citet{Wu2022} estimated fast-axis orientations and the magnitude of anisotropy in the region via adjoint travel-time tomography, and reported depth-dependent changes in fast-axis orientations. West of the SAF, the fast-axis directions reported by \citet{Porter2011} differ from those of \citet{Wu2022}, and station coverage in this region is sparse in \citet{Porter2011}. This provides a suitable example for evaluating the estimation of crustal anisotropy from enhanced virtual RFs and assessing the consistency of these estimates with the fault geometry and seismogenic processes of the region.
We observe that the anisotropy parameters derived from radial virtual receiver functions are more consistent and reliable than those obtained from linearly averaged receiver functions. Specifically, the fast axis at most stations is oriented perpendicular to the faults, indicating mineral-induced anisotropy in the lower crust.

Our approach requires fewer assumptions regarding Earth models and nuisance statistics. It generates high-quality RFs with improved backazimuth and epicentral-distance coverage, and, like other deep learning techniques, is automated and scalable across datasets and geographic regions.
The paper is structured as follows. \hyperref[sec:methodology]{Section \ref*{sec:methodology}} describes the mathematical modeling of receiver functions and condition vectors, explains the conditional diffusion transformer, and details the procedure for generating and averaging virtual RFs. \hyperref[sec:synex]{Section \ref*{sec:synex}} presents synthetic experiments evaluating the quality of virtual RFs against conventional averaging methods. \hyperref[sec:realdata]{Section \ref*{sec:realdata}} applies our method to real data from the Cascadia subduction zone and southern California, demonstrating its performance in complex geological settings. \hyperref[sec:discussion]{Section \ref*{sec:discussion}} discusses the physical consistency of virtual RFs, key assumptions, and future extensions. Appendices provide detailed mathematical derivations and implementation details.

\section{Methodology}
\label{sec:methodology}

In this section, we describe the physical basis for conditioning, the statistical properties of RFs and nuisance effects.
We develop a conditional diffusion transformer framework to generate high-quality virtual receiver functions by learning the conditional distribution of RFs given parameters associated with the incident planewave and receiver location.
The key insight is that RFs from earthquakes with similar backazimuths and epicentral distances share coherent crustal effects while exhibiting variable, earthquake-specific nuisance effects.

 \subsection{Receiver functions}
The wavefield due to a teleseismic earthquake at source position $\mathbf{x}_s$, measuring at a receiver position $\mathbf{x}_r$, can be expressed as a convolution of the earthquake source time function with the Earth's impulse response.
\begin{equation}
  \dz(t,\mathbf{x}_s, \mathbf{x}_r) = \int_{\tau}s(t-\tau)\gz(\tau,\mathbf{x}_s, \mathbf{x}_r)\,\text{d}\tau+\nz(t), 
  \label{eqn:one}
\end{equation}
\begin{equation}
    \dr(t,\mathbf{x}_s, \mathbf{x}_r) = \int_{\tau}s(t-\tau)\gr(\tau,\mathbf{x}_s, \mathbf{x}_r)\,\text{d}\tau+\mathit{\nr(t)}.
\end{equation}
Here, time is indicated as $t$, the vertical and radial components of the measured seismogram are indicated by using $\dz$ and $\dr$, respectively, while $\nz$ and $\nr$ represent uncorrelated noise in these components. We used $\gz$ and $\gr$ to denote the vertical and radial impulse responses, respectively.
The generation of radial receiver function involves the deconvolution of $\dr$ with $\dz$ to predominantly extract SV converted waves~\citep{Tauzin2019}. 
As shown in Appendix~\ref{sec:appn1}, the radial receiver function can be written as
\begin{equation}
    r(t,\mathbf{x}_s, \mathbf{x}_r)= \int_{\tau}s_{\text{a}}(t-\tau,\mathbf{x}_s, \mathbf{x}_r)\grz(\tau,\mathbf{x}_s, \mathbf{x}_r)\,\text{d}\tau + \nt(t,\mathbf{x}_s, \mathbf{x}_r), 
    \label{eqn:rfn}
\end{equation}
where $s_{\text{a}}$ denotes a zero-phase signal convolved with $\grz$, which depicts the cross-correlation between $\gr$ and $\gz$.
Reliable imaging of crustal structures using RFs relies on the accurate extraction of the crustal impulse response $\grz$, which is sensitive to the geometry and anisotropy of the crustal layers.
The interpretation of the RF is complicated by the presence of nuisance effects, which are unique to each earthquake and can be attributed to the additive term 
$\nt$. This term arises from the cross-correlation between seismic noise ($\nr$ and $\nz$) and the crustal impulse responses ($\gz$ and $\gr$). 
We use boldface symbols to denote vectors after time discretization. Accordingly, the RF in Eq.~\ref{eqn:rfn} 
is expressed in time domain after discretization as
\begin{equation}
    \rec[\n]= \sum_{\tau}\mathbf{s}_{\text{a}}[\n-\tau]\mathbf{g}^{r,z}[\tau] + \qv[\n],
    \label{eqn:rfn2}
\end{equation}
where we model the discretized receiver function $\rec \in \mathbb{R}^{\Nn_t}$, discretized impulse response $\mathbf{g}^{r,z}$, discretized zero-phase signal $\mathbf{s}_{\text{a}}$ and discretized nuisance effects $\qv$ as random vectors.
The RF $\rec$ is generated from the convolution of $\mathbf{g}^{r,z}$ and $\mathbf{s}_{\text{a}}$, with an additive noise term $\qv$.
Each sample of the random vector $\rec$ corresponds to a particular earthquake and receiver position.
We used square brackets, such as $[\n]$, to index vectors and $\Nn_t$ represents the total number of time samples. We normalize all the RFs so that the amplitude of the P wave at zero time lag is 1.

\subsection{Coherent crustal effects and nuisance effects}
\label{sec:sym}
The crustal impulse response $\mathbf{g}^{r,z}$ for a given receiver is a function of the incident teleseismic planewave, which is mostly determined by the backazimuth and epicentral distance of the earthquake.
We introduce a condition vector $\mathbf{c}$ that uniquely specifies the incident planewave propagation direction and receiver location. Specifically, $\mathbf{c}$ is a random vector that encodes the backazimuth, epicentral distance, and receiver position. 
Let us examine how the receiver function varies with the condition vector $\mathbf{c}$.
These changes can be due to the vectors $\mathbf{g}^{r,z}$, $\mathbf{s}_{\text{a}}$, or $\qv$.
Our primary assumption is that the vector $\mathbf{g}^{r,z}$ changes deterministically with $\mathbf{c}$. 
This means that for a fixed condition vector $\mathbf{c}$, the crustal impulse response $\mathbf{g}^{r,z}$  is essentially constant across different earthquakes within the corresponding backazimuth-epicentral distance bin, since $\mathbf{g}^{r,z}$ depends only on the incoming teleseismic planewave and crustal structure, not on the earthquake source.
%
This implies that $\mathbf{g}^{r,z}$, which is determined by the geometry and anisotropy of the crustal layers, is stable over the spatial scale of the bin defined by $\mathbf{c}$. In other words, the crustal effects represented by $\mathbf{g}^{r,z}$ are coherent across earthquakes with similar backazimuth and epicentral distance.
Appendix.~\ref{sec:coherency}  
presents a synthetic example where the changes in bandlimited $\mathbf{g}^{r,z}$ with respect to backazimuth and epicentral distance are depicted. 

In contrast to $\mathbf{g}^{r,z}$, the vectors $\mathbf{s}_{\text{a}}$ and $\qv$ fluctuate across different earthquakes, even for a fixed $\mathbf{c}$. This is because $\mathbf{s}_{\text{a}}$ is influenced by the earthquake source time function and the propagation effects from the source to the base of the crust, which can vary significantly between earthquakes. Similarly, $\qv$ is influenced by seismic noise, which can also vary between earthquakes and stations. 
The difference in the statistical properties of $\mathbf{g}^{r,z}$, $\mathbf{s}_{\text{a}}$, and $\qv$ with respect to $\mathbf{c}$ is the key to disentangling the crustal effects from the nuisance effects in the receiver function.

More formally, coherency of crustal effects implies that the conditional density $p(\mathbf{r} \mid \mathbf{c})$ exhibits anisotropic covariance structure. Because $\mathbf{g}^{r,z}$ is nearly constant for fixed $\mathbf{c}$ while $\mathbf{s}_{\text{a}}$ and $\qv$ vary randomly, the conditional distribution forms an elongated ellipsoid in RF space: compressed (low variance) along crustal-response directions where RFs are similar, and stretched (high variance) along nuisance directions where RFs differ. Mathematically, the eigenvalues of the conditional covariance matrix 
$\mathrm{Cov}(\mathbf{r} \mid \mathbf{c})$ are small along crustal-response directions and large along nuisance directions.
We rigorously demonstrate this anisotropic property in Appendix~\ref{sec:appn2}. Figure~\ref{fig:diffusion} provides a geometric illustration: samples of RFs from a common condition are tightly clustered (low spread) along crustal directions but widely dispersed (high spread) along nuisance directions. 

The traditional approach averages observed RFs with similar $\mathbf{c}$, which means that it approximates the mean of the conditional distribution $p(\mathbf{r} \mid \mathbf{c})$. This approach has limitations:
\begin{inparaenum}
    \item it assumes that the sample mean converges to the true mean of the distribution, which is not the case when there are insufficient earthquakes in the region associated with $\mathbf{c}$;
    \item the sample mean is sensitive to outliers, which can occur when there are large nuisance effects in some RFs;
    \item when the earthquake samples have low SNR the variance along nuisance directions is large, which means that the sample mean can be significantly influenced by nuisance effects, leading to a biased estimate of crustal effects.
\end{inparaenum}
In the next section, we describe how the averaging of multiple RFs generated by a conditional diffusion model can effectively suppress nuisance effects while preserving crustal effects, even when the number of observed RFs is limited and the SNR is low.

\begin{figure}
    \centering
    \includegraphics[scale=0.7]{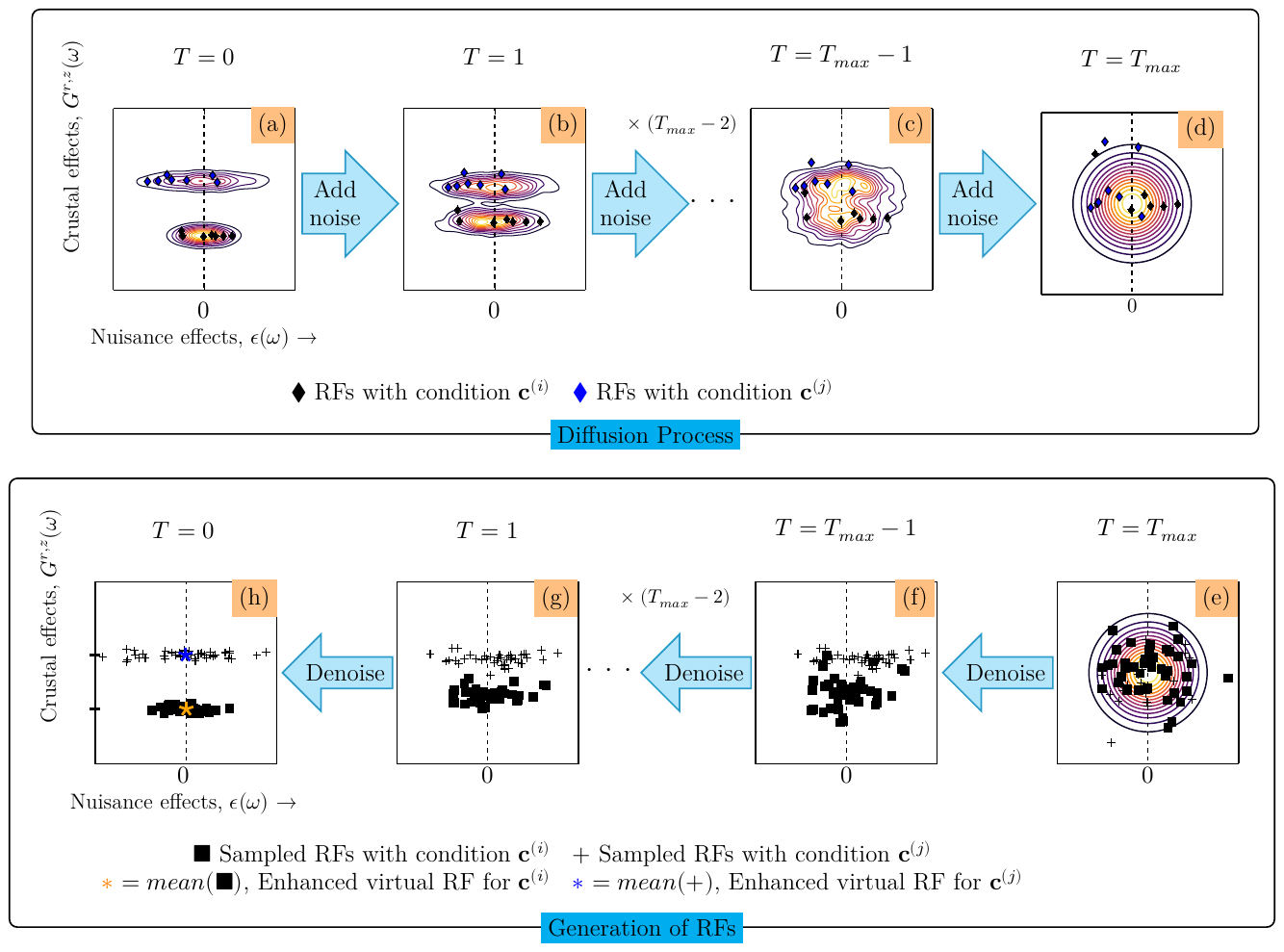}
    \caption{
        A schematic illustration of the diffusion-based generation of virtual RFs. The top box depicts the forward diffusion process, where the joint probability density function (pdf) of crustal and nuisance effects evolves over diffusion time steps $T$. 
        The bottom box illustrates the generation process, where the trained diffusion model transforms random samples from a Gaussian pdf into realistic RFs conditioned on either $\mathbf{c}^{(i)}$ or $\mathbf{c}^{(j)}$.
        (a) Measured RFs for two conditions $\mathbf{c}^{(i)}$ and $\mathbf{c}^{(j)}$ (e.g., two different backazimuths) are shown as blue and black diamond markers. 
        (b) and (c) The forward diffusion process gradually adds noise to the RFs in (a) leading to a joint pdf that becomes increasingly blurred and eventually converges to a Gaussian distribution at large $T$ as shown in (d).
(e) During the generation process, we start with random samples drawn from the Gaussian distribution, and either condition using $\mathbf{c}^{(i)}$ (plus markers) or $\mathbf{c}^{(j)}$ (square markers), to gradually transform these samples into realistic RFs as shown in (f) and (g).
(h) The generated RFs exhibit high variance along nuisance directions and low variance along crustal-effect directions, reflecting the  structure in (a). By averaging these generated RFs for a specific condition, we obtain virtual RFs with reduced nuisance effects (plotted using star markers).
}
        
      \label{fig:diffusion}
\end{figure}

\subsection{Conditional diffusion model for generating virtual receiver functions}
Learning the probability distribution $p(\mathbf{r})$ from observed samples is the essence of generative modeling. The goal is to optimize network parameters $\theta$ such that samples drawn from the learned distribution $p_\theta(\mathbf{r})$ closely resemble the observed samples. Diffusion models are a particularly effective class of generative models that have demonstrated superior performance in learning complex, high-dimensional distributions.
Diffusion models learn to generate data by reversing a gradual noising process (see Fig.~\ref{fig:diffusion}). The forward process maps each RF training example through a series of latent variables by repeatedly blending the current representation with random noise. After sufficient steps, the representation becomes indistinguishable from white noise. Since these steps are small, the reverse denoising process at each step can be approximated with a normal distribution and predicted by a deep learning model. The loss function is based on the evidence lower bound (ELBO) and ultimately results in a simple least-squares formulation that trains the network to predict the noise added at each step. 
%

%



The diffusion modeling framework naturally extends to conditional generation: by conditioning the denoising network on $\mathbf{c}$, the model learns to generate RFs that respect the crustal structure implied by the specified backazimuth, epicentral distance, and receiver position. 
%
For a more detailed mathematical formulation of the conditional diffusion model, including details on the forward and reverse processes, we refer the reader to Appendix~\ref{sec:diffusion_details}.
The conditional diffusion model learns this anisotropic structure from the training data --- i.e., it learns that RFs with similar $\mathbf{c}$ cluster tightly along crustal directions and disperse widely along nuisance directions.
This means that when the trained model generates samples with a fixed condition vector, the samples exhibit high variance along nuisance directions (reflecting earthquake source and noise variability) and low variance along crustal-effect directions (reflecting stable incoming planewave).

In this paper, 
we approximate the true conditional distribution $p(\mathbf{r} \mid \mathbf{c})$ of receiver functions given condition vectors by training a conditional diffusion transformer, which is a type of neural network architecture that combines the strengths of diffusion models and transformers (see Appendix~\ref{sec:diffusion_details} for details). The learned model is given by $p_\theta(\mathbf{r} \mid \mathbf{c})$, where $\theta$ denotes the transformer network parameters learned during training.
Finally, note that the anisotropic structure of the conditional distribution $p(\mathbf{r} \mid \mathbf{c})$ is learned implicitly from the training data and enables the diffusion model to disentangle crustal structure from nuisance effects. 
While the current discussion focuses on the radial RFs derived from the P-wave, the same framework can be applied to transverse RFs and RFs derived from other seismic phases.


\subsubsection{Training datapoints and condition vectors}

We organize and label the training data by condition vectors. Each receiver function serves as a single datapoint in the training dataset. The $i$-th training sample is denoted as $(\mathbf{r}^{(i)}, \mathbf{c}^{(i)})$, where $\mathbf{r}^{(i)}$ is the receiver function and $\mathbf{c}^{(i)}$ is its corresponding condition vector:
\begin{equation}
    \mathbf{c}^{(i)} = (\Delta^{(i)}, \phi^{(i)}, \mathbf{x}_r^{(i)}),
\end{equation}
which encodes the epicentral distance $\Delta^{(i)}$, backazimuth $\phi^{(i)}$, and receiver position $\mathbf{x}_r^{(i)}$ of the earthquake that generated it. The training dataset $\{(\mathbf{r}^{(i)}, \mathbf{c}^{(i)})\}_{i=1}^{N}$ aggregates all individual RFs from all earthquakes and all stations, where $N$ is the total number of training samples. Importantly, RFs sharing similar condition vectors (i.e., from a common backazimuth-epicentral distance bin at a given station) have coherent crustal effects but random, earthquake-specific nuisance effects. 
The parameters of the conditional diffusion transformer are optimized by minimizing the loss function (see Appendix~\ref{sec:diffusion_details} for architecture details) over this training dataset.
The training procedure is applied to both radial and transverse RF components, separately.

\subsubsection{Virtual RF generation via conditional sampling and averaging}

\paragraph{Conditional sampling for fixed crustal effects.}

After training, the learned model $p_\theta(\mathbf{r} \mid \mathbf{c})$ can generate virtual RFs for any specified condition vector $\mathbf{c}$. Consider the $i$-th training condition $\mathbf{c}^{(i)} = (\Delta^{(i)}, \phi^{(i)}, \mathbf{x}_r^{(i)})$. We can draw multiple independent virtual RF samples from the conditional distribution:
\begin{equation}
    \vir^{(k)} \sim p_\theta(\mathbf{r} \mid \mathbf{c} = \mathbf{c}^{(i)}), \quad k = 1, 2, \ldots, K.
\end{equation}
Each sample $\vir^{(k)}$ exhibits different nuisance effects.
Because the conditional covariance is anisotropic (Section~\ref{sec:sym}), these samples remain tightly clustered in crustal-response directions while exhibiting variability in nuisance-dominated directions. That is, all samples $\vir^{(k)}$ share the same coherent crustal effects determined by $\mathbf{c}^{(i)}$ (i.e., the incoming planewave), but each realization exhibits different nuisance effects arising from earthquake source variability and random noise. This sampling process, depicted in Figure~\ref{fig:diffusion}, effectively generates virtual RFs with controlled crustal structure but realistic variability in nuisance terms.
In the supplementary material, we plotted the generated samples $\vir^{(k)}$ for a fixed condition $\mathbf{c}^{(i)}$.

\paragraph{Flexible generation: interpolation.}
After training, the learned model $p_\theta(\mathbf{r} \mid \mathbf{c})$ can generate virtual RFs that interpolate across the entire range of backazimuths, epicentral distances, represented in the training data for a given receiver. This means that we can generate virtual RFs for any condition vector $\mathbf{c}$, even those not present in the training dataset.
Let us say we have training conditions $\mathbf{c}^{(i)}$ and $\mathbf{c}^{(j)}$ corresponding to backazimuth-epicentral distance bins at a given station. We can generate virtual RFs for an intermediate condition $\mathbf{c}^{(k)}$ that lies between $\mathbf{c}^{(i)}$ and $\mathbf{c}^{(j)}$ by sampling from $p_\theta(\mathbf{r} \mid \mathbf{c} = \mathbf{c}^{(k)})$. The generated virtual RFs for $\mathbf{c}^{(k)}$ might exhibit crustal effects that are consistent with the incoming teleseismic planewave defined by $\mathbf{c}^{(k)}$, even though this specific condition was not present in the training data.
Similarly, we can generate virtual RFs for intermediate receiver positions between two stations in the training dataset.
Note that interpolation does not add new information, but it allows us to better interpret RFs and their variations across backazimuth and epicentral distance, especially in regions with sparse earthquake coverage.
 We rely on the network's capacity for high-dimensional nonlinear interpolation~\citep{hornik1989multilayer} to generate realistic RFs at intermediate backazimuths, epicentral distances, or even receiver positions not present in the training dataset. This interpolation capability allows us to fill gaps in earthquake coverage and produce densely sampled virtual RFs across the full range of backazimuths and epicentral distances.
The interpolations are not guaranteed to be physically accurate, but we validate them by comparing the generated virtual RFs with true RFs for earthquakes that were held out of the training dataset. We demonstrate this validation in our synthetic experiments (Section~\ref{sec:synex}).



\paragraph{Averaging to enhance signal-to-noise ratio.}

For a given condition $\mathbf{c}^{(i)}$, to obtain a virtual RF with enhanced quality, we average the $K$ samples:
\begin{equation}
    \hat{\vir}(\mathbf{c}^{(i)}) = \frac{1}{K} \sum_{k=1}^{K} \vir^{(k)}.
    \label{eqn:virtualRF}
\end{equation}
Note that $\mathbf{c}^{(i)}$ can be any condition vector, including those corresponding to training conditions or interpolated conditions.
Why does averaging suppress nuisance while preserving crustal effects? Because the samples $\vir^{(k)}$ are drawn from an anisotropic distribution, they cluster tightly along crustal directions but vary widely along nuisance directions. When we average, the consistent crustal components reinforce each other (constructive interference), while the variable nuisance components cancel out (destructive interference). Mathematically, the averaged virtual RF approximates the mean of the conditional distribution $p(\mathbf r\mid\mathbf c)$, which enhances the expected crustal response for the specified condition as the nuisance term $\qv$ has zero conditional mean 
(see Appendix~\ref{sec:appn2}, where we show that $\mathbb{E}[\qv \mid \mathbf{c}] = 0$).
This means that as $K$ increases, the averaged virtual RF $\hat{\vir}(\mathbf{c}^{(i)})$ converges to the crustal response for the specified condition, while the influence of nuisance effects diminishes.
Thus the averaged virtual RF isolates the deterministic crustal response while averaging away earthquake source and nuisance variability.

\section{Synthetic experiment}
\label{sec:synex}

\begin{figure}
    \centering
    \includegraphics[trim={0cm 0.5cm 0 0.2cm},scale=0.9]{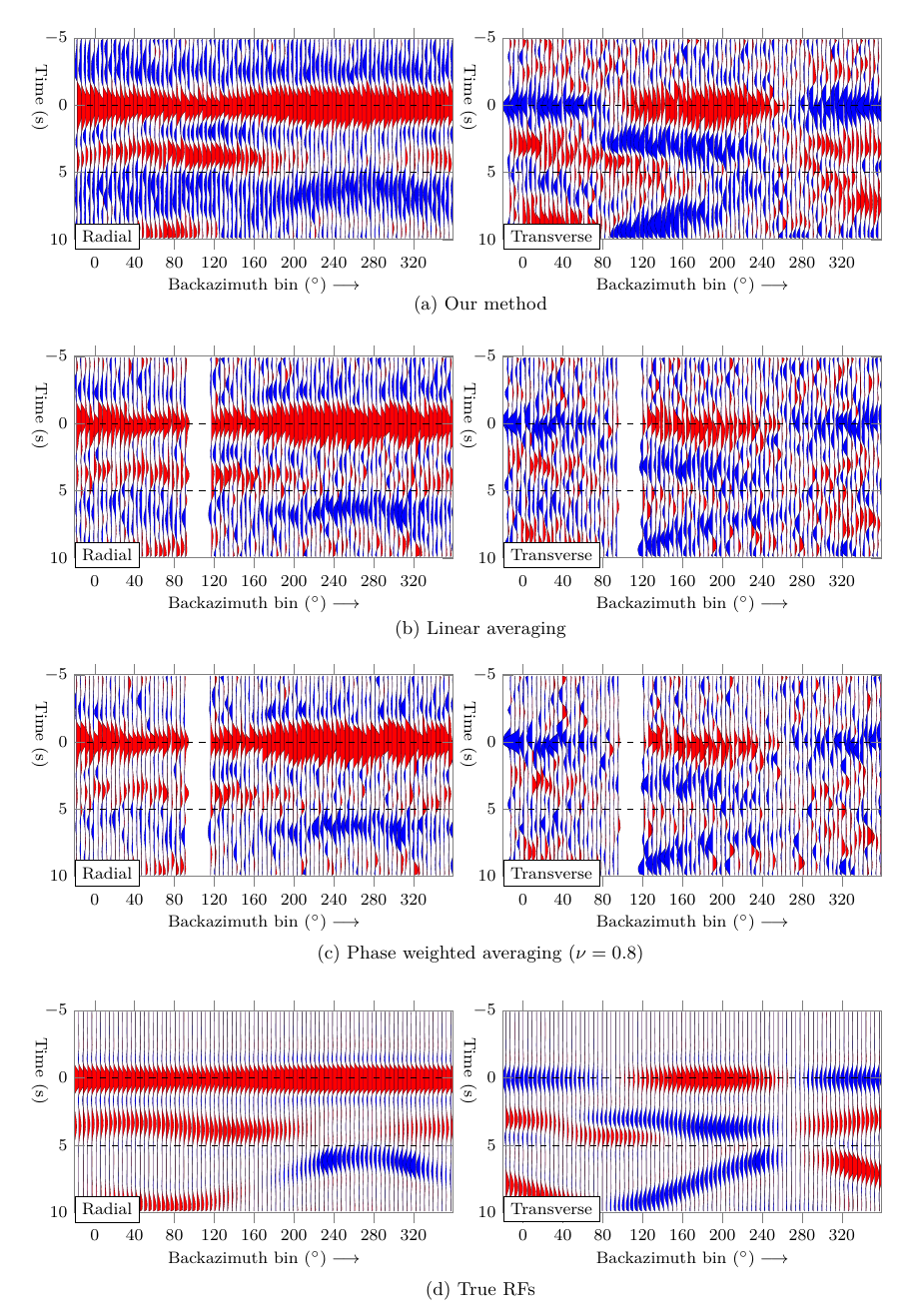}
    \vspace{0.8cm}
    \caption{Radial and transverse RFs from the synthetic experiment, derived using (a) diffusion transformer (b) linear bin-wise averaging and 
    (c) phase-weighted averaging, are compared to (d) true RFs.
    A station with a dipping $30$km thick crustal layer is considered.
    }
\label{fig:model1}
\end{figure}

\begin{figure}
\centering
\includegraphics[width=\textwidth]{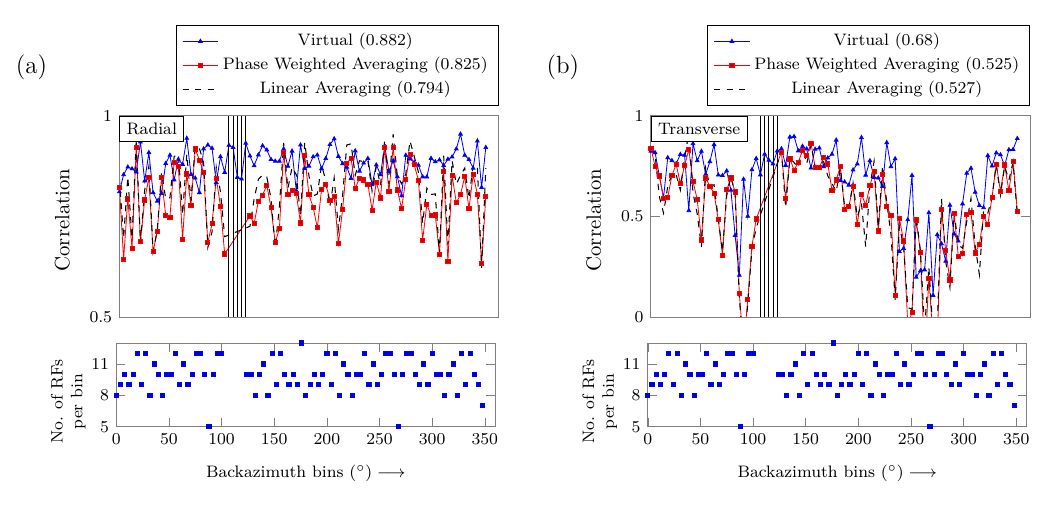}
\caption{Enhancement of RFs for the synthetic experiment, where the quality of the (a) radial and (b) transverse RFs, plotted in Fig.~\ref{fig:model1}, is assessed using the normalized correlation coefficient.
On average, the normalized correlation coefficient of the generated virtual RFs is higher than that of linearly averaged and phase-weighted RFs, indicating higher correlation with the true RFs.
The backazimuth distribution of the synthetic dataset is realistic, with a  highlighted gap between $100^{\circ}$ and $120^{\circ}$.
\label{fig:mse1}}
\end{figure}
To validate our method, we designed a synthetic experiment that tests three critical capabilities: (1) {nuisance suppression}---whether virtual RFs exhibit reduced noise compared to traditional averaging methods, (2) {interpolation accuracy}---whether the model can generate physically realistic RFs in regions with missing earthquake coverage (backazimuth gaps), and (3) {preservation of crustal structure}---whether subtle crustal features such as anisotropy and layer geometry are retained in the virtual RFs. We constructed a synthetic dataset with known crustal structure, realistic non-Gaussian noise derived from ambient seismic recordings, and a deliberate backazimuth gap ($100^{\circ}$ to $120^{\circ}$) to test interpolation capabilities. This experiment enables quantitative assessment of virtual RF quality through direct comparison with noise-free true RFs.

%
The crustal velocity model used in this synthetic experiment consists of a 30\,km-thick crust over an isotropic mantle. The crust has $8\%$ azimuthal anisotropy with a fast-axis direction of $60^{\circ}$ N. The crust--mantle interface dips by $20^{\circ}$ N from the horizontal with a strike of $0^{\circ}$. We use crustal properties of $V_p=6.0$ km/s, $V_s=3.45$ km/s, and density $2.8$ g/cc, and mantle properties of $V_p=8.04$ km/s, $V_s=4.47$ km/s, and density $3.6$ g/cc.
This anisotropy and dip produce systematic, backazimuth-dependent variations in the converted phases originating at crust–mantle interfaces. Synthetic RFs for a range of backazimuths and epicentral distances are computed from this crustal model following the steps outlined below.
\paragraph*{1. Crustal impulse responses.}
The crustal impulse responses for the vertical, radial, and transverse components were calculated using the PyRaysum software~\citep{Bloch_Audet_2023}. They were computed for all backazimuths from $0^{\circ}$ to $360^{\circ}$, except for the range $100^{\circ}$ to $120^{\circ}$, with an interval of $2^{\circ}$. The epicentral distances range from $30^{\circ}$ to $95^{\circ}$, with a spacing of $1^{\circ}$.
As shown in Fig.~\ref{fig:mse1}, the non-uniformity comes from event counts per backazimuth--epicentral-distance bin (some bins contain more earthquakes than others), which mimics realistic earthquake occurrence.
%


\paragraph*{2. Earthquake signatures.}
To create synthetic seismograms, the components of each crustal impulse response were convolved with a unique signature of the earthquake source obtained from real seismograms, after trimming a 40\,s window centered around the P-wave arrival. This approach not only mimics the bandwidth variations and zeros found in the source spectrum typical of real earthquakes but also confirms
that there are no repeated earthquake occurrences.


\paragraph*{3. Adding ambient seismic noise.}
Ambient noise from seismic stations in the US was added to all three components independently. The noise amplitude was then scaled to achieve a target signal-to-noise ratio (SNR), with SNR values randomly drawn from a truncated log-normal distribution between 0.1 and 1.
We calculated the SNR by calculating the ratio of the root-mean-squared amplitude $10$\,s before and after the P-wave arrival on the radial component seismogram.

\paragraph*{4. Deconvolution and training data.}
Frequency domain deconvolution was employed to generate synthetic radial and transverse RFs.
The deconvolution process was regularized using a water level parameter of $0.01$ and a Gaussian filter with a width of $5$. 
The backazimuth, epicentral distance, and station location for each synthetic RF were stored in the waveform headers and later used as conditioning parameters. We set the latitude and longitude of the synthetic station to $0^{\circ}$. For validation, 10\% of all RFs were randomly selected, and the remaining RFs were used as training data for the diffusion transformer.




After training the network, we generated virtual RFs for conditions where backazimuth spans from $0^{\circ}$ to $360^{\circ}$ at $4^{\circ}$ interval, including those within the backazimuth gap, using Eq.~\ref{eqn:virtualRF} with $K=40$. For these conditions, the epicentral distance was fixed at $50^{\circ}$. We also tested the performance of traditional averaging, i.e., we calculated both linearly averaged and phase-weighted averaged RFs for the same set of backazimuths, using a bin size of $4^{\circ}$. Figure~\ref{fig:model1} shows these RFs for both the radial and transverse components. 
%
To assess the accuracy of the RFs, we used true RFs obtained with the same steps as above but employed a Gaussian source wavelet and omitted the ambient noise addition.
The true RFs are plotted in Fig.~\ref{fig:model1}d.
The virtual RFs generated by the diffusion model exhibit reduced nuisance effects compared
to linear and phase-weighted averaging (power = $0.8$), plotted in Figs.~\ref{fig:model1}b and \ref{fig:model1}c, respectively.
This is clear because the normalized correlation coefficient (NCC) between the virtual RFs and true RFs exceeds that of linear or phase-weighted averaging, as shown in Fig.~\ref{fig:mse1}. In this plot, the NCC values of the virtual radial RFs remain largely unaffected by the number of earthquakes (or RFs) in each bin, unlike the averaged radial RFs, whose NCC values vary with bins. In contrast, the NCCs of the transverse RFs exhibit greater variability with backazimuth, as expected, because the energy of converted seismic phases in the transverse component is very small at certain backazimuths. Nevertheless, the virtual transverse RFs show a systematically higher correlation with the true RFs than with the averaged RFs. The change in polarity along the backazimuth is also clearly observed in the transverse virtual RFs. 
As shown in Fig.~\ref{fig:mse1}a, the virtual RFs generated within the backazimuth gap ($100^{\circ}$ to $120^{\circ}$) correlate well with the true RFs. This indicates that the diffusion model successfully learns patterns from the training data and produces RFs that are physically meaningful and consistent, even in regions with limited coverage.
%



\section{Results using real data}
\label{sec:realdata}

Having validated our method on synthetic data with known ground truth, we now apply it to real receiver functions from two geologically complex settings. These settings serve complementary purposes: the Cascadia subduction zone tests whether virtual RFs can resolve fine-scale velocity contrasts in a steeply dipping, multi-layered structure (the subducting oceanic plate), while southern California tests whether virtual RFs enable robust estimation of crustal anisotropy parameters in a tectonically active transform margin. Both regions present challenges for traditional RF analysis: sparse and non-uniform earthquake distributions, low signal-to-noise ratios at many stations, and complex crustal structure that produces subtle, backazimuth-dependent variations requiring careful comparison across multiple earthquakes. Our results demonstrate that receiver functions obtained with our method enhance the interpretation and enable more reliable estimation of crustal anisotropy.

\begin{figure}
\centering
\includegraphics[width=\textwidth]{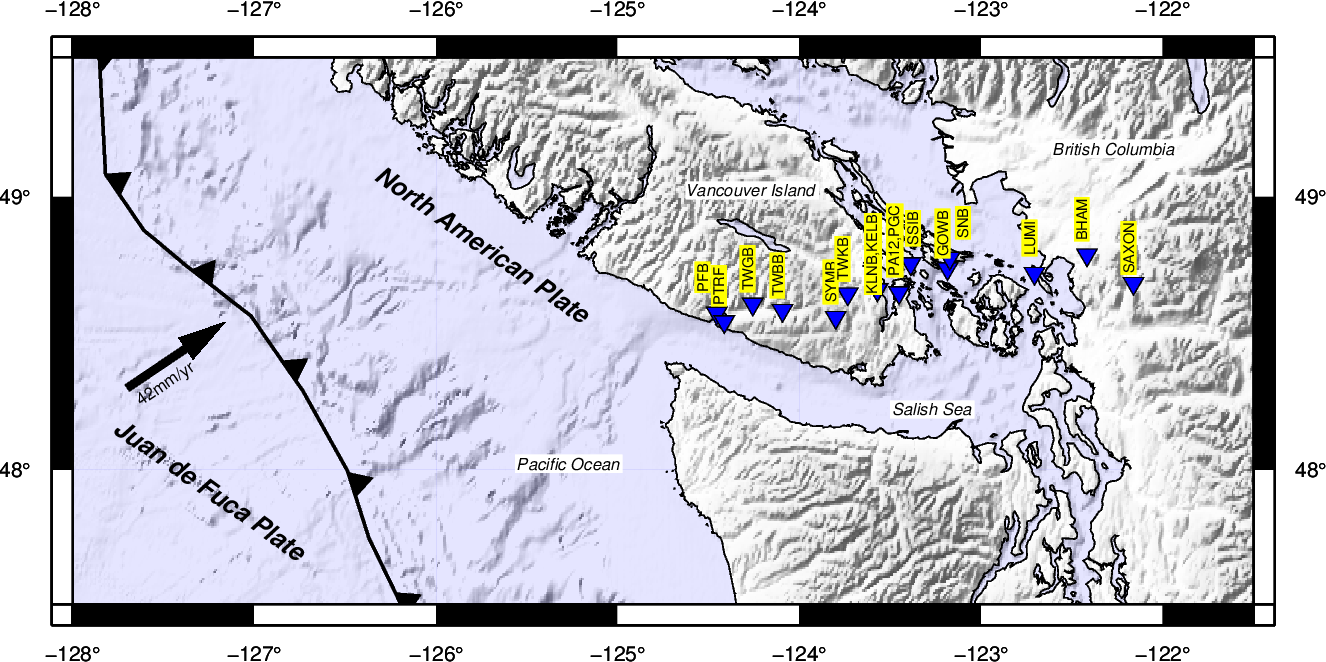}
\caption{Map illustrating the tectonic setting and positions of the stations (represented by black and white triangles) utilized in the study. This map also depicts the subduction of the Juan de Fuca plate beneath the North American plate, occurring at a rate of $42$ mm/yr ~\citep{DeMets1994} (indicated by a black arrow). The solid black line adorned with triangles marks the subduction boundary. Stations are from the POLARIS ~\citep{POL}, C8 ~\citep{c8}, CN ~\citep{cn}, and UW ~\citep{uw} networks. This paper plots RFs from stations marked by blue inverted triangles that are approximately positioned along a transect. 
}
\label{fig:casmap}
\end{figure}

\paragraph*{Data preparation and training.}
For the Cascadia subduction zone, we downloaded teleseismic earthquakes recorded by $16$ seismic stations on southern Vancouver Island, British Columbia and the Olympic Peninsula, belonging to the POLARIS~\citep{POL}, C8~\citep{c8}, CN~\citep{cn} and UW~\citep{uw} networks were used. These seismic stations, shown in Fig.~\ref{fig:casmap}, cover the Vancouver Island segment of the Cascadia Subduction Zone. 
In the case of Southern California, we used teleseismic data from $80$ seismic stations belonging to CI network~\citep{Caltech1926} (see Fig.~\ref{fig:calianiso}). 
In both cases,
earthquakes with epicentral distances ranging from $30^{\circ}$ to $100^{\circ}$ and a minimum moment magnitude of $5.5$ were used. The seismograms were rotated to radial and transverse components, then bandpass filtered using $0.03$ -- $3.0$ Hz, and trimmed at $20$\,s before the PREM-calculated P arrival time and $50$\,s after.
Radial and transverse RFs were generated using water-level deconvolution with a Gaussian factor of $5.0$ and a water level of $0.01$. For validation, 10\% of all RFs were randomly selected, and the remaining RFs were used as training data for the diffusion transformer. RFs from Cascadia and southern California were trained separately.


\paragraph*{Linear averaging procedure for real data.}
For comparison with our method, we also calculated linearly averaged RFs employing a systematic binning procedure. For each region (Cascadia and Southern California), we organized all available RFs into two-dimensional bins defined by backazimuth and epicentral distance ranges. Specifically, backazimuth bins were defined with $4^{\circ}$ intervals (e.g., $0^{\circ}$--$4^{\circ}$, $4^{\circ}$--$8^{\circ}$, etc.), and epicentral distance bins were defined with $5^{\circ}$ intervals (e.g., $30^{\circ}$--$35^{\circ}$, $35^{\circ}$--$40^{\circ}$, etc.). For each bin, all RFs falling within the backazimuth and epicentral distance range were arithmetically averaged to produce a single representative RF. This approach assumes that RFs within a bin have similar crustal structure (a key assumption validated in Appendix~\ref{sec:coherency}), and the averaging suppresses uncorrelated noise and earthquake-specific nuisance effects. However, this method has inherent limitations that were discussed earlier. In the rest of this section, we compare the virtual RFs generated by our diffusion-based method to these linearly averaged RFs, where we generated virtual RFs for conditions with uniformly distributed backazimuths and epicentral distances for the stations, according to Eq.~\ref{eqn:virtualRF} with $K=50$. 

\begin{figure}
\centering
\includegraphics[width=\textwidth]{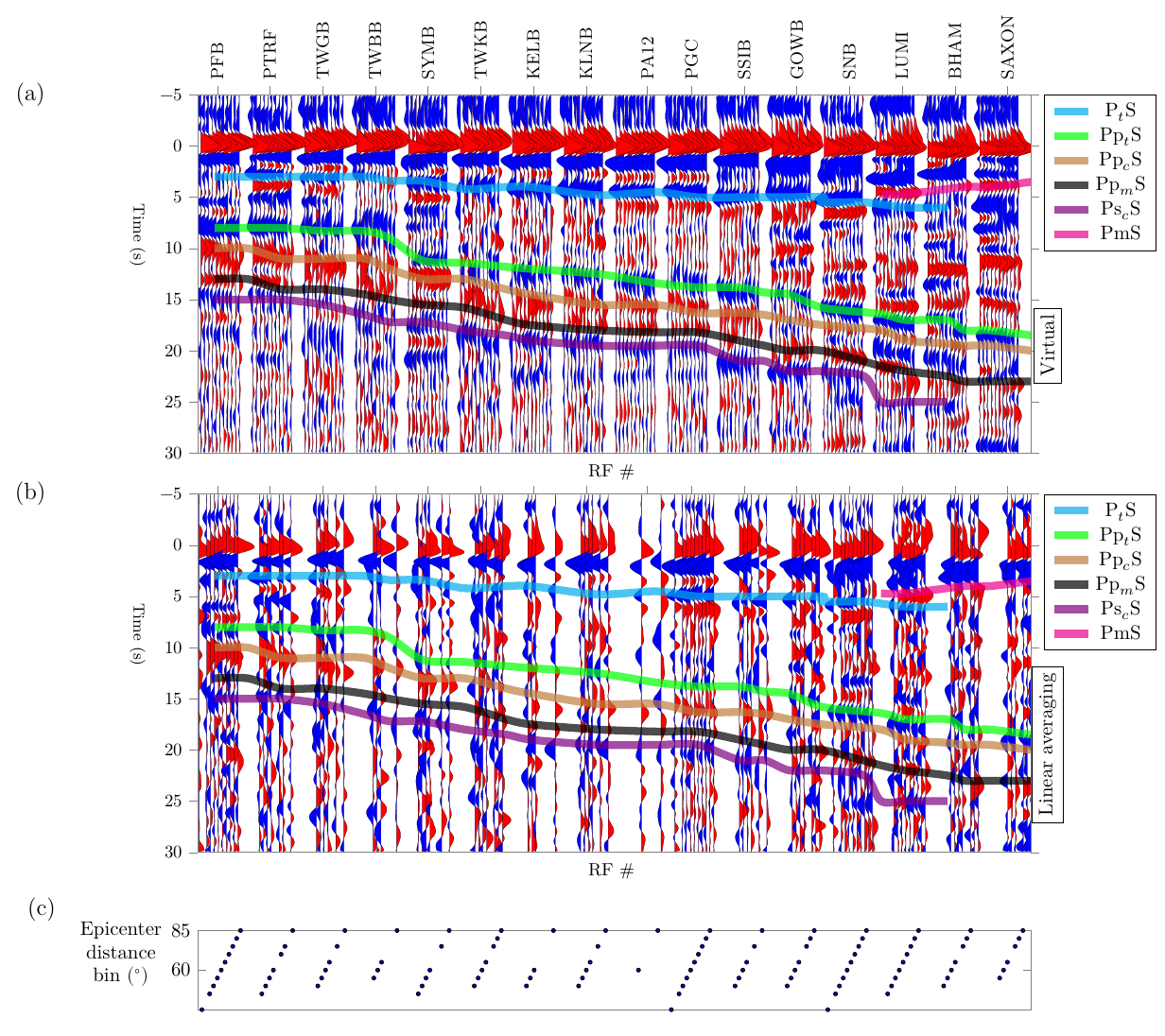}
\caption{(a) Virtual and (b) linearly averaged radial RFs at seismic stations (Fig.~\ref{fig:casmap}) traversing the Cascadia subduction zone. The RFs in (b) are computed for the backazimuth bin centered at $300^{\circ}$. Panel (c) shows the epicentral-distance bin corresponding to (b). The virtual RFs in (a) span epicentral distances from $35^{\circ}$ to $85^{\circ}$, with the backazimuth fixed at $300^{\circ}$. Key seismic phases from the subducted slab (highlighted by colored lines) are more clearly resolved in the virtual RFs than in the linearly averaged RFs. The phase notation follows ~\cite{Bloch2023}, as listed in Tab.~\ref{tab:phases_notation}}.
\label{fig:casrad}
\end{figure}
 
\begin{figure}
\centering
\includegraphics[width=\textwidth]{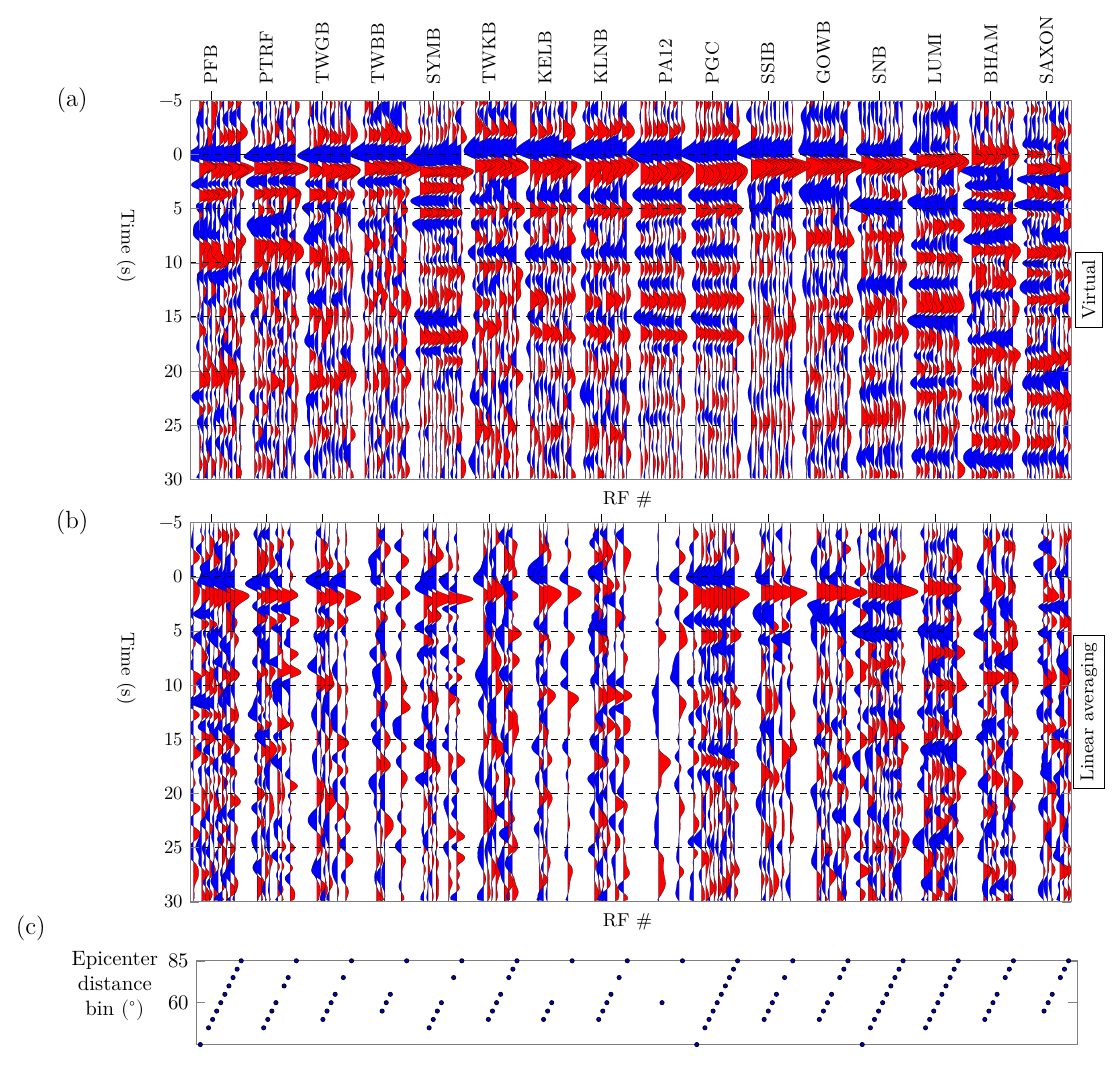}
\caption{Similar to Fig.~\ref{fig:casrad}, but pertaining to transverse RFs. It is important to observe that the virtual RFs exhibit greater complexity and enhancement, allowing for the converted phases to be smoothly traced along the transect.}
\label{fig:castran}
\end{figure}

\subsection{Cascadia subduction zone}
 Studies by \cite{Bloch2023} and \cite{Audet2009Seismic} characterize the subducting oceanic slab as two dipping layers: a low-velocity zone overlying a higher-velocity zone. This structure produces three seismic velocity interfaces, comprising a negative velocity contrast at the top of the slab and two positive velocity contrasts beneath. When upgoing P waves encounter these interfaces, they generate multiple scattered S waves (detailed in Table~\ref{tab:phases_notation}). Conventional linear averaging within backazimuth and epicentral-distance bins often leaves gaps in spatial coverage. Our method seeks to more clearly image the scattered phases associated with the subducting slab.
 Figs.~\ref{fig:casrad} and \ref{fig:castran} show radial and transverse RFs for stations that transect the Cascadia subduction zone, respectively. For each station, RFs are sorted according to increasing epicentral distance. 
For virtual RFs, the epicentral distance spans from $35^{\circ}$ to $85^{\circ}$ in steps of $5^{\circ}$, with the backazimuth fixed at $300^{\circ}$. 
For linear averaging, each RF is associated with the available epicentral distance bins, while the backazimuth bin is kept constant at $300^{\circ}$. We used RFs recorded from the same backazimuth to remove variations caused by anisotropy, thereby enhancing the comparability of RFs between stations. The virtual RFs produced from our method are in agreement with those of ~\cite{Bloch2023} and ~\cite{Audet2009Seismic} but with substantially improved RF coverage (see supplementary material). Compared with linearly averaged radial RFs, virtual radial RFs are more consistent across stations. The scattered seismic phases in virtual radial RFs, highlighted by colored lines, are observed at nearly all stations. In time windows where these scattered seismic phases are absent, RF amplitudes should be close to zero if nuisance effects are effectively minimized. In the virtual radial RFs, at times later than these seismic phases, the RF amplitudes are indeed small compared to those during the seismic phases. This is particularly clear at stations PFB and PTRF, where the amplitudes in the $20$--$30$ s time window are very small relative to the seismic-phase time window ($5$--$15$ s). In contrast, for linearly averaged radial RFs, the amplitudes in this window are similar in magnitude to those observed within the seismic-phase window. This indicates that nuisance effects are effectively reduced in the virtual RFs but not in the linearly averaged RFs. Similarly, virtual transverse RFs in Fig.~\ref{fig:castran} show a marked improvement in resolving crustal effects, which is difficult to achieve with conventional receiver function techniques due to the low signal-to-noise ratio of the transverse component.  To examine the variation of RFs with backazimuth, we also plotted the RFs for station SNB as a function of backazimuth; the figure is provided in the supplementary material. Virtual RFs provide additional backazimuth coverage that is not accessible with linearly averaged RFs. The polarity reversal of the transverse RFs with backazimuth is clearly revealed by the virtual RFs at SNB.
We will now discuss the key seismic phases observed in the virtual radial RFs (Fig.~\ref{fig:casrad}) in more detail and compare them to those observed in the linearly averaged RFs.

\paragraph*{Forward-scattered S-waves from the subducting slab.}
The P\textsubscript{t}S seismic phase, which is forward-scattered from the top of the subducting slab, is not well resolved across stations in
previous studies~\cite{Bloch2023} --- this is evident in the linearly averaged RFs in Fig.~\ref{fig:casrad}b, where this phase from the slab, highlighted by the cyan line, is not clearly observed at many stations (e.g., TWKB, MCGB, KELB and KLNB). In contrast, this phase is clearly observed across stations in virtual radial RFs (Fig.~\ref{fig:casrad}a). The travel time of this phase gradually increases along the transect, indicating a northeast-dipping Juan de Fuca plate beneath the North American plate.
In addition, we observed a strong positive seismic phase in the virtual RFs at stations LUMI, BHAM, and SAXON between $0$ and $5$ s. We interpret this phase as the forward-scattered S-wave from the continental Moho (denoted as PmS). 

%



\paragraph*{Conversions of surface-reflected P and S-waves.}
According to~\cite{Bloch2023}, Pp\textsubscript{t}S and Pp\textsubscript{c}S are S-waves converted at the top interface, center interface, and oceanic Moho, respectively, due to P-waves reflected from the surface. These seismic phases, highlighted in Fig.~\ref{fig:casrad}, are clearly resolved and consistent across stations in the virtual RFs, except at TWGB, TWBB, and TWKB, where they are weak. In the linearly averaged RFs, although these phases are weakly present at PFB, PGC, SNB, and LUMI, they lack consistency across stations. The Ps\textsubscript{c}S phase, an S-wave converted at the center interface due to the surface-reflected S-wave, is also observed across all stations in the virtual RFs, but is absent or poorly resolved in the linearly averaged RFs.
The delay times of these seismic phases gradually increase along the station profile, indicating a slab that dipped eastward. The observation of these phases beneath station SAXON in virtual RFs suggests that the slab penetrates approximately 250 km inland from the trench. Previous RF studies did not clearly image the slab to this extent \citep{Bloch2023,Calvert2020}.

In summary, the virtual RFs obtained with our method exhibit minimal nuisance effects, allowing a much clearer delineation of scattered S-waves from the multilayered subducting slab than is achieved with linearly averaged RFs.  This shows that our method performs well in complex geological environments and enables a reliable interpretation of RFs.

\begin{table}
  \centering
  \begin{tabular}{|c|c|}
    \hline
    Phase & Notation \\ \hline
    Forward scattered S wave from top velocity contrast & P\textsubscript{t}S \\ \hline
    S wave from top velocity contrast due to surface-reflected S wave & Ps\textsubscript{t}S \\ \hline
    S wave from top velocity contrast due to surface-reflected P wave & Pp\textsubscript{t}S \\ \hline
    S wave from middle velocity contrast due to surface-reflected P wave  & Pp\textsubscript{c}S \\ \hline
    S wave from bottom velocity contrast due to surface-reflected P wave & Pp\textsubscript{m}S \\ \hline
  \end{tabular}
  \vspace{1cm}
  \caption{This table illustrates the phases and their respective symbols as referenced in ~\cite{Bloch2023}. According to ~\cite{Bloch2023}, the subducting slab beneath the southern Vancouver Island is conceptualized with two distinct layers: one being a low-velocity layer resting above the oceanic crust. This configuration causes a negative velocity contrast at the top interface and two positive velocity contrasts: one occurs between the low-velocity layer and the oceanic crust, and the other occurs between the oceanic crust and the mantle.}
  \label{tab:phases_notation}
\end{table}



\subsection{Southern California}

\begin{figure}
\centering
\includegraphics[width=\textwidth]{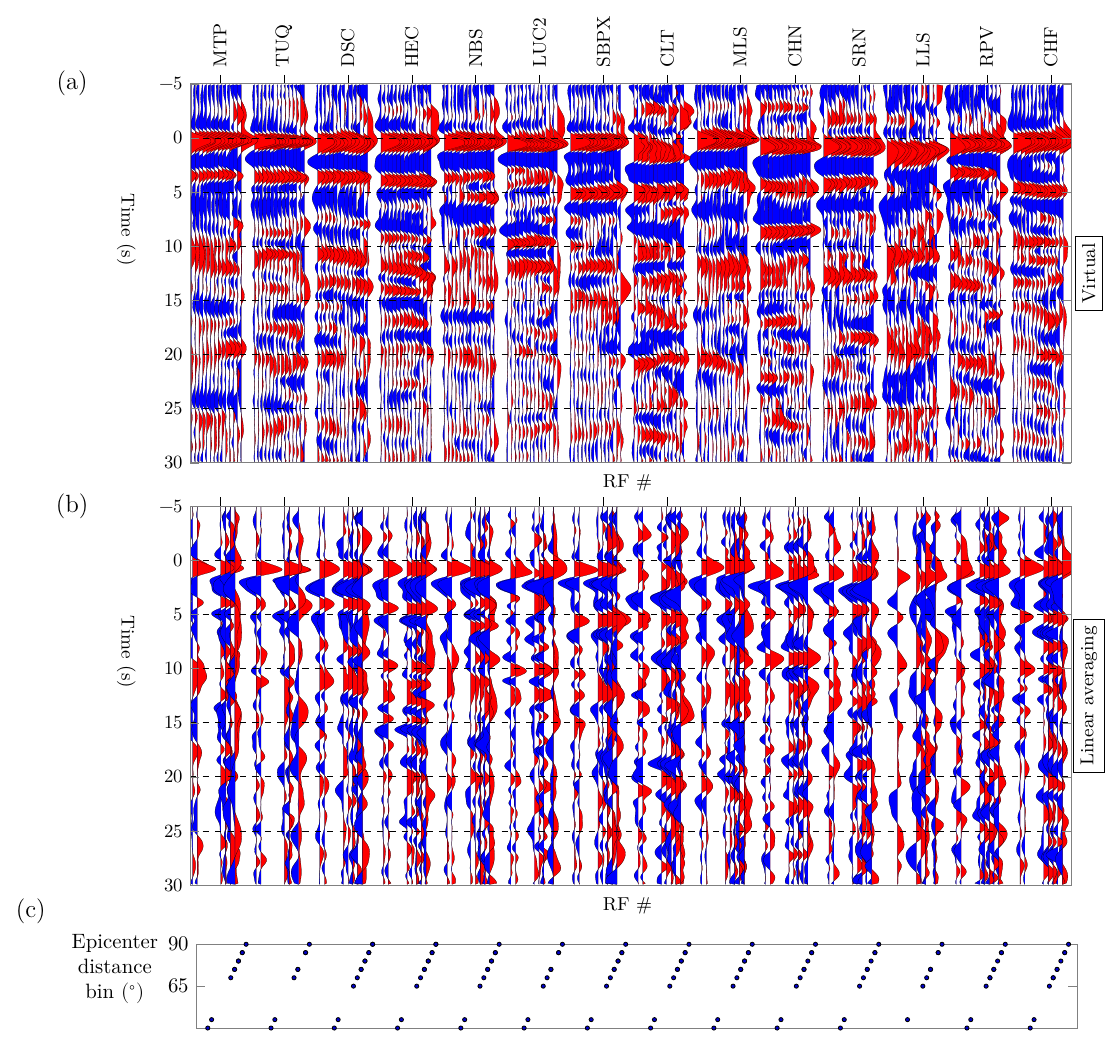}
\caption{(a) Virtual and (b) linearly averaged radial RFs for a station profile transecting the San Andreas Fault (SAF), San Jacinto Fault Zone (SJFZ) and Elsinore Fault (EF) in Southern California. Station locations are given in supplementary material. At each station, RFs from different epicentral distance are plotted. Backazimuth angle is fixed at $304^{\circ}$. (c) shows epicentral distance bin corresponding to (b) linearly averaged RFs. Positive amplitude at around 5 s, interpreted as converted S-wave from Moho, shows the variation of Moho depth along the profile. Virtual RFs resemble linearly averaged RFs but demonstrate increased quality across bins and stations.}
\label{fig:calirad}
\end{figure}

 Southern California has a complex network of active faults that mark the boundary between the Pacific and North American plates. The main fault lines include the San Andreas Fault (SAF), the San Jacinto Fault Zone (SJFZ), and the Elsinore Fault (EF), all of which run nearly parallel to the coast. 
In Fig.~\ref{fig:calirad}, the virtual radial RFs derived from our method reveal significant undulations in Moho depth along the profile crossing the SAF, SJFZ, and EF. The highest Moho depths are observed beneath stations SBPX and CLT, consistent with previous studies~\citep{zhu2000moho,Ozakin2015}. RFs from the transverse component are given in the supplementary material. Compared with simple linear stacking (Fig.~\ref{fig:calirad}b), the virtual RFs exhibit clearer seismic phases and improved epicentral distance coverage. Anisotropy studies in this region have also revealed a complex pattern of azimuthal anisotropy directions around the San Andreas Fault (SAF) ~\citep{Porter2011}, as well as depth-dependent variations ~\citep{Wu2022}.
Crustal azimuthal anisotropy arises primarily due to the preferred alignment of minerals and the orientation of stress-induced microcracks. Platy or elongated minerals, such as micas or amphiboles, can become aligned by tectonic deformation, causing seismic waves to travel faster parallel to this fabric than perpendicular to it. The axis parallel to the fabric is referred to as the fast axis. Similarly, differential stress can open or orient microcracks along specific directions, resulting in anisotropy. Together, these processes imprint a record of past and present deformation in the crust, observable through seismic anisotropy measurements. 

\begin{figure}
\centering
\includegraphics[width=1\textwidth]{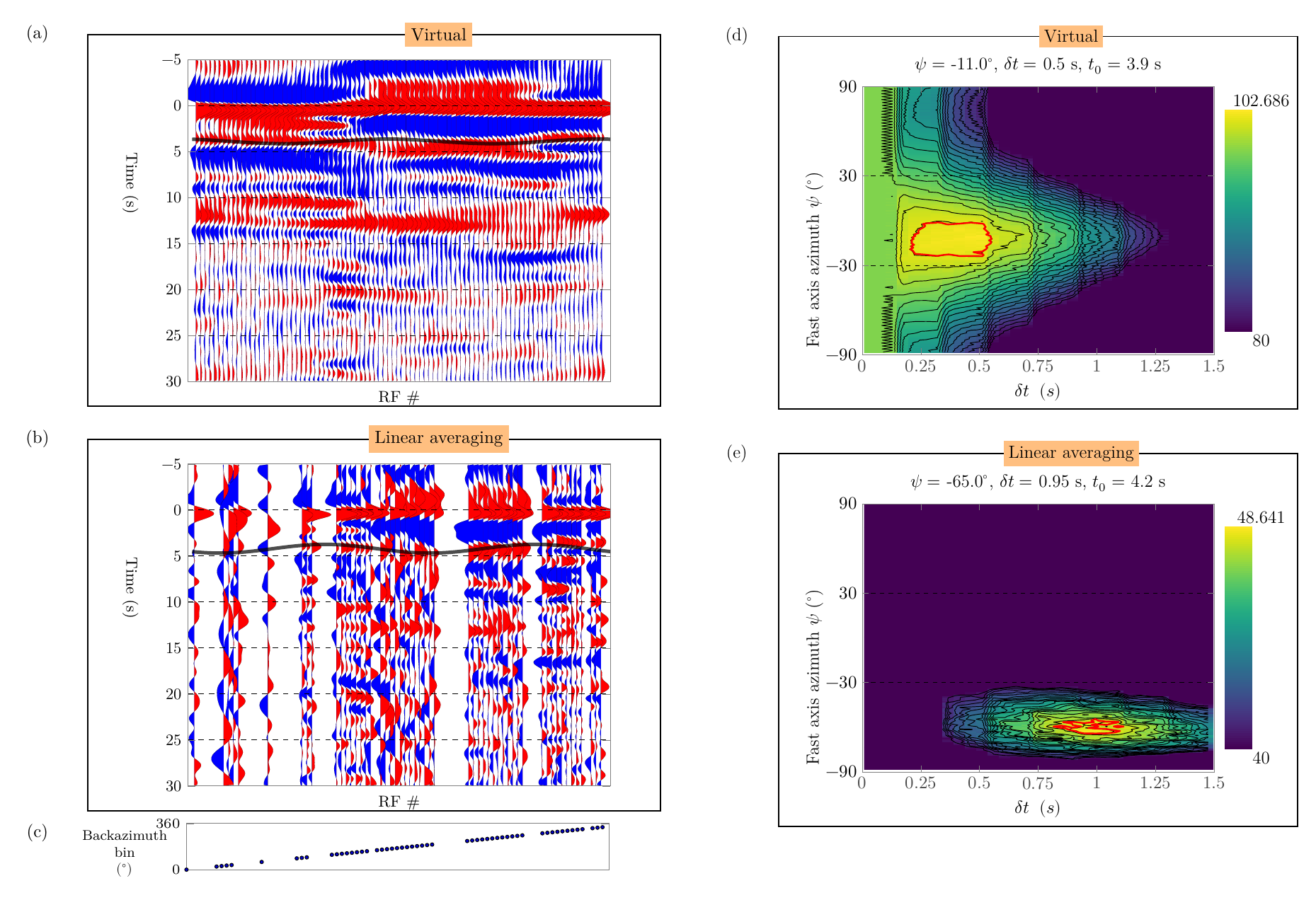}
\caption{This figure shows (a) virtual RFs generated from the diffusion transformer and (b) linearly averaged RFs at seismic station NBS in southern California (location is given in Fig.~\ref{fig:calianiso}). Panel (c) shows the backazimuth bins corresponding to the RFs in (b). Panels (d) and (e) show estimated fast-axis azimuth ($\fastbz$), S-wave splitting time ($\delt$) and isotropic travel time ($t_{0}$) from (a) and (b) respectively. The travel time of converted S-wave corresponding to the estimated anisotropy parameters is indicated by solid black line in (a) and (b). The full backazimuth coverage of the virtual RFs leads to improved parameter estimation, as indicated by the evenly spaced contours. The red lines in (d) and (e) denote the 98\% confidence regions.}
\label{fig:heatmap}
\end{figure}

Various approaches have been employed to estimate crustal azimuthal anisotropy from RFs, including forward modeling \citep{Porter2011,Ozacar2009}, harmonic decomposition \cite{Audet2015}, and cosine fitting \cite{liu2015crustal,Zheng2018}. The reliability of these methods depends critically on both the quality and backazimuthal coverage of the RFs as discussed above.
%
%
In this paper, we estimated the fast-axis azimuth $\fastbz$ and delay time $\delt$ (travel time difference between the fast S-wave and slow S-wave) from the virtual radial RFs in southern California using the cosine-fitting technique (detailed in Appendix ~\ref{sec:anisotropy_method}). To evaluate the performance of virtual RFs, we also estimated the anisotropy from linearly averaged RFs using the same technique. Fig.~\ref{fig:heatmap} presents the virtual radial RFs and the linearly averaged RFs for the station NBS, along with their estimated $\fastbz$ and $\delt$. The distribution of fitness across different combinations of $\fastbz$ and $\delt$ at the chosen $t_{0}$ is shown as a heatmap. The heatmap in Figure~\ref{fig:heatmap}d shows the fitness distribution pertaining to the virtual RFs, which attains its maximum at $\fastbz = -11^{\circ}$, $\delt = 0.5$ s, and $t_{0} = 3.9$ s. While the heatmap in Figure~\ref{fig:heatmap}e presents the corresponding distribution for linearly averaged RFs, with a maximum at $\fastbz = -65^{\circ}$, $\delt = 0.95$ s, and $t_{0} = 4.2$ s. The fitness distribution corresponding to the virtual RFs exhibits smoother, more regularly spaced contours than that from the linearly averaged RFs, indicating that the parameters ($\fastbz$ and $\delt$) are better constrained and estimated with higher reliability. The elongated contours in Fig.~\ref{fig:heatmap}e demonstrate a trade-off between the two parameters, which is consistent with the non-uniform backazimuthal coverage evident in the linearly averaged RFs (Fig.~\ref{fig:heatmap}b), in contrast to the uniform coverage of the virtual RFs. The supplementary material provides a corresponding figure for the DSC station, analogous to Fig.~\ref{fig:heatmap}. These results indicate that the anisotropy parameters ($\fastbz$ and $\delt$) inferred from the virtual RFs are more reliable than those obtained from the linearly averaged RFs. Next, we present the estimated anisotropy for seismic stations in southern California and discuss their implications.

\begin{figure}
\centering
\includegraphics[width=0.8\textwidth]{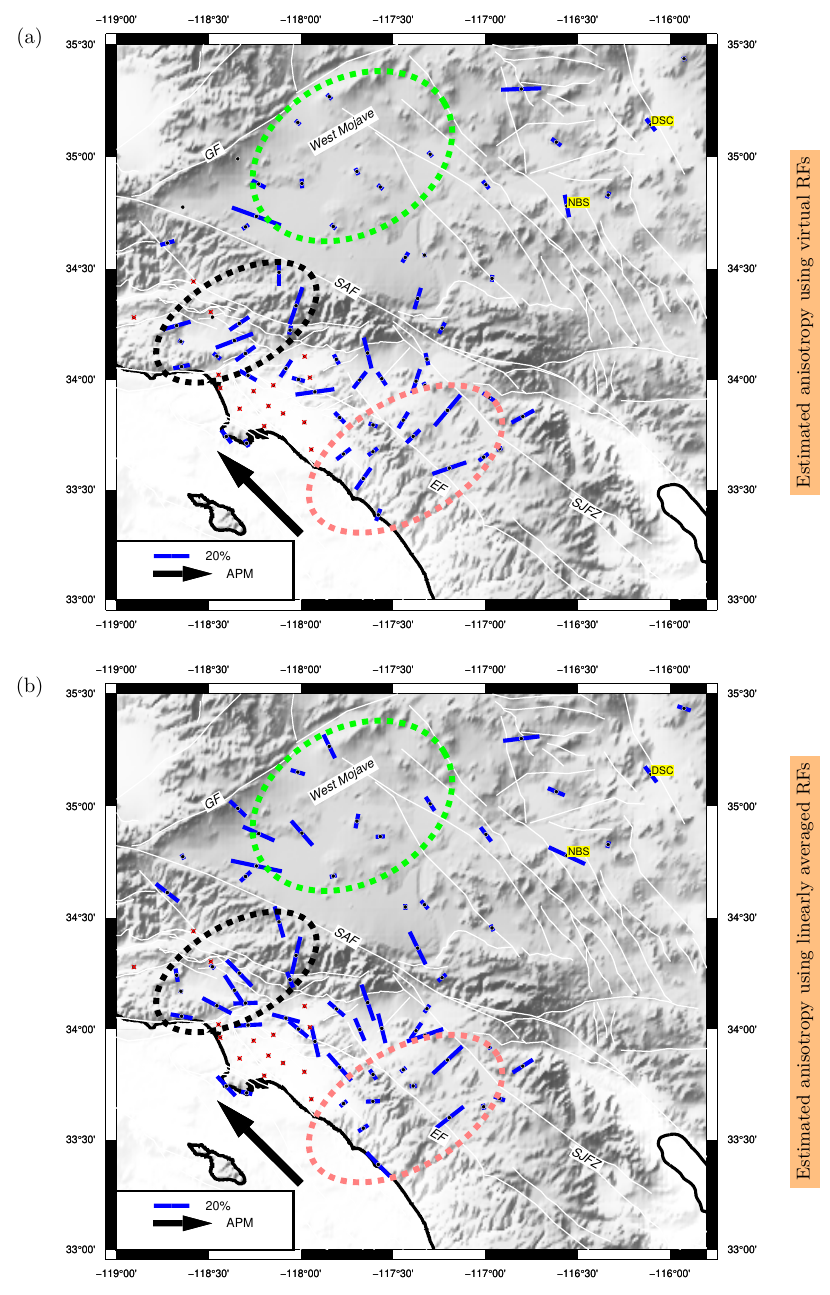}
\caption{The map shows southern California with estimated seismic anisotropy plotted at the locations of the individual stations. The azimuth of each blue solid line indicates the fast-axis direction, and its length is proportional to the anisotropy percentage. Panels (a) and (b) display results from virtual receiver functions and linearly averaged RFs, respectively. The black arrow denotes the absolute plate motion of the Pacific plate. White solid lines mark regional faults, with the San Andreas Fault (SAF), San Jacinto Fault Zone (SJFZ), Garlock Fault (GF), and Elsinore Fault (EF) highlighted. Black and red dashed ellipses outline the Central Transverse Ranges (CTR) and Western Peninsular Ranges (WPR), respectively, and the green dashed ellipse marks the western Mojave region. Red cross indicates noisy seismic stations which we excluded from the anisotropy estimation.
}
\label{fig:calianiso}
\end{figure}

Figure~\ref{fig:calianiso} presents the fast-axis direction and anisotropy percentage across southern California, estimated from both virtual and linearly averaged RFs. Here, the anisotropy percentage represents the magnitude of anisotropy and is defined as the relative delay time, $\delta t$, given as a percentage of the isotropic travel time $t_0$. Only stations whose RF section plots exhibit a clear P-wave arrival at zero time lag and a distinct forward-scattered S-wave are included in the anisotropy estimates. Stations that do not meet these criteria are excluded and marked with red crosses. Anisotropy estimates based on virtual RFs exhibit higher spatial similarity within local regions than those obtained from linearly averaged RFs. We identified three areas where the estimates are spatially similar, and these regions are discussed below.

\paragraph*{Western Mojave --- small anisotropy consistent with crustal stress-field}  Anisotropy estimation using virtual RFs shows a small anisotropy percentage in the western Mojave, highlighted by the green dashed ellipse in Fig.~\ref{fig:calianiso}a, which agrees with the P-wave anisotropy results of \citet{Wu2022}. The crustal stress field analysis by \citet{Yang2016} also indicates smaller stress magnitudes in this region compared with the surrounding areas. For linearly averaged RFs, the estimated percentage of anisotropy in this region is comparatively larger, which can be attributed to the biased estimation discussed above. 

\paragraph*{Central Transverse Ranges --- fault-perpendicular anisotropy consistent with P-wave anisotropy results.}
In Fig.~\ref{fig:calianiso}a, the Central Transverse Ranges (CTR), highlighted by the black dashed ellipse, show an anisotropy direction that is nearly perpendicular to the SAF. This fault-perpendicular anisotropy is also reported by \citet{Wu2022GRL}. In contrast, estimates from linearly averaged RFs (Fig.~\ref{fig:calianiso}b) in the same region are inconsistent and do not agree with \citet{Wu2022GRL}.

\paragraph*{Western Peninsular Ranges --- fault-perpendicular anisotropy inconsistent with P-wave anisotropy results.}
The Western Peninsular Ranges (WPR), highlighted by the red dashed ellipse in Fig.~\ref{fig:calianiso}a, also exhibit an anisotropy direction that is perpendicular to the San Jacinto Fault (SJF) and the Elsinore Fault (EF). \citet{Wu2022GRL} likewise reported pronounced anisotropy, but with a fault-parallel orientation, which they linked to an old imprint of previous Farallon plate subduction. 
This discrepancy between our results and those of \citet{Wu2022GRL} may arise from the fact that the anisotropy estimated from P-waves of local earthquakes is more sensitive to shallower structures, while the anisotropy derived from RFs is sensitive to the entire crustal column.
%
Estimates from linearly averaged RFs (Fig.~\ref{fig:calianiso}b) show weak anisotropy in this region, which contrasts with \citet{Wu2022GRL}.

In summary, the virtual RF–based results indicate stronger anisotropy west of the SAF than to the east, consistent with the present transform faulting regime: the Pacific plate crust (west of the SAF) is more highly stressed, whereas the North American plate crust (east of the SAF) experiences comparatively lower stress. East of the SAF, our estimates reveal fault-perpendicular anisotropy in two regions, WPR and CTR. A comprehensive examination of how these anisotropy measurements relate to the underlying tectonic processes is beyond the scope of this study. Nevertheless, robust estimation of anisotropy using virtual RFs can improve interpretations and support future studies.


\paragraph*{}

\subsection{Sanity checks}

Because virtual RFs are generated by a neural network rather than directly measured, we must verify that they represent genuine crustal structure rather than spurious patterns, overfitting artifacts, or unphysical interpolations. We perform three physical consistency tests designed to catch common failure modes: (1) \textit{spatial coherence}---virtual RFs from nearby stations (see, for example, RFs for PFB and PTRF, and KELB and KLNB in Cascadia) should exhibit similar crustal signatures where geological continuity is expected, validating that the model learns true crustal structure rather than station-specific noise; (2) \textit{smooth interpolation}---virtual RFs generated at intermediate backazimuths and epicentral distances should vary smoothly and consistently with neighboring conditions, confirming that the network performs physically meaningful interpolation rather than erratic extrapolation; (3) \textit{gap validity}---virtual RFs generated within earthquake coverage gaps should resemble patterns from adjacent conditions, demonstrating that the model generalizes appropriately rather than inventing arbitrary structure. Together, these checks ensure that the diffusion model extracts stable, geologically meaningful crustal responses.

\section{Discussion}
\label{sec:discussion}

This research assumes that the RFs from nearby earthquakes share coherent crustal effects.
Coherency implies minimal traveltime differences for converted arrivals from such earthquakes. What constitutes "minimal" and "nearby" depends on the maximum frequency of interest and the depth of the target structure on the receiver side.
For example, to successfully extract converted seismic waves from the crust or upper mantle, it is possible to group relatively distant earthquakes.
As travel time differences increase for deeper mantle structures, the earthquakes must be nearer.

The proposed method effectively reduces nuisance effects in RFs and enhances their spatial coverage, offering substantial potential for improving RF analysis and interpretation.
It leverages multiple RFs simultaneously and reducing nuisance effects with a unified network, it does not process each RF independently, compared to most of the existing methods.
In other words, it learns the structure of the RF data and the distribution of nuisance effects across multiple RFs, rather than treating each RF separately. 

Training the diffusion transformer, similar to any deep neural network, necessitates hyperparameter tuning (e.g., model depth, hidden dimension, number of heads, and noise schedule). Selecting these optimally improves generative performance. In our case, hyperparameter tuning is carried out through synthetic experiments.
Finally, several promising directions remain unexplored: (1) joint training on multiple seismic phases (P, S, SKS, PKP) to leverage complementary sampling of crustal structure, 
 (2) uncertainty quantification through analysis of the variance of generated samples, and (3) extension to deeper mantle structures with appropriate bin size adjustments.

\section{Conclusions}
\label{sec:conclusions}
Receiver functions (RFs) are an essential technique for analyzing crust and mantle structures beneath seismic stations; however, they are often contaminated with nuisance effects that reduce the precision and reliability of RF measurements. We introduce a conditional diffusion transformer to reduce these nuisance effects by learning the distribution of RFs and generating virtual RFs through conditional sampling and averaging. Our approach outperforms linear and phase-weighted averaging in denoising synthetic RFs affected by realistic non-Gaussian nuisance effects, resulting in consistent RFs. Virtual RFs clearly depict polarity reversals in transverse RFs along the backazimuth, distinguishing them from averaged RFs. Radial RFs from seismic stations in the Cascadia Subduction Zone forearc reveal scattered S-waves originating from the subducting slab and show strong agreement with previous studies. 
Crustal azimuthal anisotropy estimation requires RFs with minimized nuisance effects and adequate spatial coverage—difficult to achieve with conventional RF processing. Conditional diffusion transformers enable us to generate realistic virtual RFs even in regions with gaps in backazimuth or epicentral distance, thereby providing full spatial coverage.
We estimated crustal azimuthal anisotropy in southern California using both virtual and linearly averaged RFs. Virtual RFs provide better-constrained anisotropy estimates that are more consistent with previous studies than those from linearly averaged RFs.
The method is robust because it integrates all available RFs regardless of signal quality, provides full automation for scalability and reproducibility, and improves RF analysis by suppressing nuisance variability. Diffusion-based conditional generation enables more precise interpretations of crustal and mantle structures and reveals insights previously obscured by noise.

\textbf{Acknowledgments}
This work is funded by the Science and Engineering Research Board, Department of Science and Technology, India (Grant Number SRG/2021/000205).
This study was enabled by Julia and Python programming languages. PyTorch package ~\citep{NEURIPS2019_9015} was used to model and train neural networks. Data were downloaded from IRIS Data Center and Natural Resources Canada Data Center using ObsPyDMT~\citep{Hosseini2017}.

\textbf{Data availability}
Earthquake waveform data used in this study are available in \href{http://service.iris.edu/}{IRIS Data Center}, \href{https://www.earthquakescanada.nrcan.gc.ca}{Natural Resources Canada Data Center} and \href{https://service.scedc.caltech.edu/}{Southern California Data Center}. All codes and packages used in this study are open-access.

\bibliographystyle{unsrtnat}
\bibliography{references_dit}

\appendix

\section{Nuisance effects in receiver functions}
\label{sec:appn1}
\setcounter{equation}{0}
\renewcommand{\theequation}{\thesection.\arabic{equation}}

In this appendix, we show that
the radial receiver function (RF) $r$ from a single teleseismic earthquake can be expressed as follows:
\begin{equation}
    r(t,\mathbf{x}_s, \mathbf{x}_r)= \int_{\tau}s_{\text{a}}(t-\tau,\mathbf{x}_s, \mathbf{x}_r)\grz(\tau,\mathbf{x}_s, \mathbf{x}_r)\text{d}\tau+ \nt(t,\mathbf{x}_s, \mathbf{x}_r).
    \label{eqn:A1}
\end{equation}
Here, $s_{\text{a}}$ represents a zero-phase signal, $\grz$ denotes the crustal impulse response, $\mathbf{x}_s$ and $\mathbf{x}_r$ are the source and receiver positions respectively, and $\nt$ accounts for noise terms. 
The radial receiver function (RF) is calculated by deconvolving the vertical component seismogram from the radial component seismogram as described in Eq. ~\ref{eqn:one}. 
Fourier transform converts the convolution in the time domain into multiplication in frequency domain. Deconvolution is the process of reversing the effects of convolution and thus can be expressed as a division in the frequency domain.
We express the measured 
seismograms in frequency domain as
\begin{equation}
\begin{split}
  \Dz(\omega,\mathbf{x}_s, \mathbf{x}_r) = S(\omega)\Gz(\omega,\mathbf{x}_s, \mathbf{x}_r)+\Nz(\omega) \quad \text{and} \\
  \Dr(\omega,\mathbf{x}_s, \mathbf{x}_r) = S(\omega)\Gr(\omega,\mathbf{x}_s, \mathbf{x}_r)+\Nr(\omega),
  \label{eqn:A2}
  \end{split}
\end{equation}
where $\omega$ is the angular frequency. Then, the radial receiver function is given by
\begin{equation}
    \label{eqn:A3}
    \mathit{R}(\omega,\mathbf{x}_s, \mathbf{x}_r) = \frac{\Dr(\omega,\mathbf{x}_s, \mathbf{x}_r)\cc{\Dz(\omega,\mathbf{x}_s, \mathbf{x}_r)}}{|\Dz(\omega,\mathbf{x}_s, \mathbf{x}_r)|^2},
 \end{equation}  
where the complex conjugate of the vertical component is $\cc{\Dz(\omega,\p)}$.
It is important to recognize that the spectrum of $\Dz$ may contain zeros or very small values, causing instability. To address this issue, various regularization techniques can be employed. We represent the regularization factor by $\lambda$, which varies according to the chosen regularization method, resulting in
\begin{equation}
    \mathit{R}(\omega,\mathbf{x}_s, \mathbf{x}_r) = \lambda(\omega)\Dr(\omega,\mathbf{x}_s, \mathbf{x}_r)\cc{\Dz(\omega,\mathbf{x}_s, \mathbf{x}_r)}.
   \label{eqn:A4}
\end{equation}
For example, 
 in the case of water-level regularization ~\citep{clayton1976source},
 \begin{equation}
 \lambda(\omega,\mathbf{x}_s, \mathbf{x}_r) = \frac{W(\omega)}{\text{max}\big(|\Dz(\omega,\mathbf{x}_s, \mathbf{x}_r)|^{2}, c\;\underset{\forall\,\omega}{\text{max}}\,(|\Dz(\omega,\mathbf{x}_s, \mathbf{x}_r)|^{2})\big)},
 \end{equation}
 where $c$ is the water-level parameter and $W(\omega)= \text{exp}\,(-\frac{\omega}{4a^{2}})$ is a Gaussian filter with a width given by the parameter $a$.

By substituting Eq. ~\ref{eqn:A2} into Eq. ~\ref{eqn:A3}, we derive 
 \begin{equation}
 \begin{split}
    R(\omega,\mathbf{x}_s, \mathbf{x}_r) = \lambda(\omega,\mathbf{x}_s, \mathbf{x}_r)\big(|S(\omega)|^{2}\;\Gr(\omega,\mathbf{x}_s, \mathbf{x}_r)\cc{\Gz(\omega,\mathbf{x}_s, \mathbf{x}_r)}+\cc{S(\omega)}\;\cc{\Gz(\omega,\mathbf{x}_s, \mathbf{x}_r)}\;\Nr(\omega) \\
    + S(\omega)\Gr(\omega,\mathbf{x}_s, \mathbf{x}_r)\cc{\Nz(\omega)}  +\Nr(\omega)\cc{\Nz(\omega)}\big).
    \label{eqn:A5}
    \end{split}
\end{equation}
 In this expression, we ignore the term $\Nr(\omega)\cc{\Nz(\omega)}$, based on the assumption that the seismic noise from the radial and vertical components is uncorrelated. The term $\Gr(\omega,\mathbf{x}_s, \mathbf{x}_r)\cc{\Gz(\omega,\mathbf{x}_s, \mathbf{x}_r)}$ denotes the radial crustal impulse response, whereas $\cc{S(\omega)}\,\cc{\Gz(\omega,\mathbf{x}_s, \mathbf{x}_r)}\,\Nr(\omega)$ and $ S(\omega)\Gr(\omega,\mathbf{x}_s, \mathbf{x}_r)\cc{\Nz(\omega)}$ account for nuisance from random noise in the seismogram. 
 These nuisance effects are relatively insignificant for high-SNR seismograms but can significantly distort the radial crustal response in low-SNR seismograms. 
 The amplification of seismic noise is influenced by the signature of the source within the regularization factor $\lambda$, which plays a role in the nuisance effects in the RF.
Rewriting Eq. ~\ref{eqn:A5} results in the following equation,
\begin{equation}
    R(\omega,\mathbf{x}_s, \mathbf{x}_r)= \underbrace{S_{\text{a}}(\omega,\mathbf{x}_s, \mathbf{x}_r)\Grz(\omega,\mathbf{x}_s, \mathbf{x}_r)}_{\text{Crustal term}}+ \underbrace{\Nt(\omega,\mathbf{x}_s, \mathbf{x}_r)}_{\text{Nuisance term}},
    \label{eqn:A6}
\end{equation}
where
\begin{equation}
\begin{split}
\Grz(\omega,\mathbf{x}_s, \mathbf{x}_r)&=\Gr(\omega,\mathbf{x}_s, \mathbf{x}_r)\cc{\Gz(\omega,\mathbf{x}_s, \mathbf{x}_r)},\nonumber \\
S_{\text{a}}(\omega,\mathbf{x}_s, \mathbf{x}_r)&=\lambda(\omega,\mathbf{x}_s, \mathbf{x}_r)\,|S(\omega)|^{2},\nonumber \\
\Nt(\omega,\mathbf{x}_s, \mathbf{x}_r)&=\lambda(\omega,\mathbf{x}_s, \mathbf{x}_r)\cc{S(\omega)}\;\cc{\Gz(\omega,\mathbf{x}_s, \mathbf{x}_r)}\;\Nr(\omega)+\lambda(\omega,\mathbf{x}_s, \mathbf{x}_r)S(\omega)\Gr(\omega,\mathbf{x}_s, \mathbf{x}_r)\cc{\Nz(\omega)} \nonumber.
\end{split}
\end{equation}
Importantly, $S_{\text{a}}(\omega,\mathbf{x}_s, \mathbf{x}_r)$ represents a zero-phase signal, influenced by both the regularization method and the source signature. In the time domain, Eq. ~\ref{eqn:A6} corresponds to Eq. ~\ref{eqn:A1}.

\section{Statistical properties of the nuisance term}
\label{sec:appn2}
\setcounter{equation}{0}
\renewcommand{\theequation}{\thesection.\arabic{equation}}
In this appendix, we analyze the statistical properties of the nuisance term $\Nt$ in the receiver function (RF) expression derived in Appendix~\ref{sec:appn1}, when conditioning on the vector 
\begin{equation}
\mathbf{c} = (\Delta, \phi, \mathbf{x}_r),
\label{eqn:B0a}
\end{equation}
where $\Delta$ is the epicentral distance, $\phi$ is the backazimuth, and $\mathbf{x}_r$ is the receiver position. We compare the statistical properties of the nuisance term with those of the structural term in the RF expression.

\subsection{Conditional mean of the nuisance term}
From the definition,
\begin{equation}
\Nt(\omega,\mathbf{x}_s, \mathbf{x}_r)=
\lambda(\omega,\mathbf{x}_s, \mathbf{x}_r)\cc{S(\omega)}\,\cc{\Gz(\omega,\mathbf{x}_s, \mathbf{x}_r)}\Nr(\omega)
+
\lambda(\omega,\mathbf{x}_s, \mathbf{x}_r)S(\omega)\Gr(\omega,\mathbf{x}_s, \mathbf{x}_r)\cc{\Nz(\omega)}.
\label{eqn:B2}
\end{equation}

We assume that the noise processes $\Nr(\omega)$ and $\Nz(\omega)$:

\begin{enumerate}
    \item have zero mean,
    \item are mutually uncorrelated,
    \item are statistically independent of the observable condition vector $\mathbf{c} = (\Delta, \phi, \mathbf{x}_r)$.
\end{enumerate}

Under these assumptions,
\begin{equation}
\mathbb{E}\big[\Nr(\omega)\big] = 
\mathbb{E}\big[\Nz(\omega)\big] = 0,
\end{equation}
which immediately implies
\begin{equation}
\mathbb{E}\big[\Nt(\omega,\mathbf{x}_s, \mathbf{x}_r)\mid \mathbf{c}\big] = 0.
\label{eqn:B3}
\end{equation}
Therefore, for every fixed condition vector $\mathbf{c} = (\Delta, \phi, \mathbf{x}_r)$, i.e., for all earthquakes with the same epicentral distance, backazimuth, and recorded at the same receiver, the nuisance term has zero mean.

\subsection{Conditional variance of the nuisance term}

Although the nuisance term has zero conditional mean, its covariance depends on $\mathbf{c}$. Defining
\[
A_r(\omega,\mathbf{x}_s, \mathbf{x}_r)=\lambda(\omega,\mathbf{x}_s, \mathbf{x}_r)\cc{S}\,\cc{\Gz(\omega,\mathbf{x}_s, \mathbf{x}_r)},
\qquad
A_z(\omega,\mathbf{x}_s, \mathbf{x}_r)=\lambda(\omega,\mathbf{x}_s, \mathbf{x}_r)S\Gr(\omega,\mathbf{x}_s, \mathbf{x}_r),
\]
Eq.~\ref{eqn:B2} can be written compactly as
\begin{equation}
\Nt(\omega,\mathbf{x}_s, \mathbf{x}_r) = A_r(\omega,\mathbf{x}_s, \mathbf{x}_r)\Nr(\omega) + A_z(\omega,\mathbf{x}_s, \mathbf{x}_r)\cc{\Nz(\omega)}.
\label{eqn:B4}
\end{equation}

Assuming
\[
\mathrm{Var}(\Nr)=\sigma_r^2, 
\qquad
\mathrm{Var}(\Nz)=\sigma_z^2,
\]
the conditional variance of nuisance terms becomes
\begin{equation}
\mathrm{Var}\big(\Nt \mid \mathbf{c}\big) = \mathbb{E}\big[|A_r(\omega,\mathbf{x}_s, \mathbf{x}_r)|^2 \mid \mathbf{c}\big] \sigma_r^2 + \mathbb{E}\big[|A_z(\omega,\mathbf{x}_s, \mathbf{x}_r)|^2 \mid \mathbf{c}\big] \sigma_z^2.
\label{eqn:B5}
\end{equation}
This means that the covariance of the terms in $\Nt$ varies with the condition vector $\mathbf{c}$.

\subsection{Comparison with the structural term}
In contrast to the nuisance terms, the crustal structure term in Eq.~\ref{eqn:A6}, has two factors: the source autocorrelation term $S_{\text{a}}(\omega,\mathbf{x}_s, \mathbf{x}_r)$ and the crustal impulse response $\Grz(\omega,\mathbf{x}_s, \mathbf{x}_r)$. The factor $\Grz$ varies deterministically with the condition vector $\mathbf{c}$, while $S_{\text{a}}$ does not vary deterministically with $\mathbf{c}$ due to its dependence on the earthquake source spectrum, which can differ among earthquakes sharing the same $\mathbf{c}$. However, both factors are zero-phase and real-valued, and we assume that the average source power spectrum across earthquakes in the bin defined by $\mathbf{c}$ is non-zero. Therefore, the conditional mean of the structural term is non-zero, which is a crucial distinction from the nuisance term that has zero conditional mean.
In other words, the conditional mean is given by
\begin{equation}
\mathbb{E}\big[S_{\text{a}}(\omega,\mathbf{x}_s, \mathbf{x}_r)\Grz(\omega,\mathbf{x}_s, \mathbf{x}_r) \mid \mathbf{c}\big] = \hat{S}_{\text{a}}(\omega, \mathbf{c})\, \Grz(\omega, \mathbf{c}) \neq 0,
\label{eqn:B6}
\end{equation}
where $\hat{S}_{\text{a}}(\omega, \mathbf{c})$ is the average source autocorrelation term across earthquakes in the bin defined by $\mathbf{c}$, and $\Grz(\omega, \mathbf{c})$ is the deterministic crustal impulse response for that bin. Note that the non-zero conditional mean of the structural term arises from the fact that both $\hat{S}_{\text{a}}(\omega, \mathbf{c})$ and $\Grz(\omega, \mathbf{c})$ are non-trivial and zero-phase, in the frequency band of interest. 
In this paper, 
using the diffusion model, we aim to approximate the conditional mean of the structural term by generating virtual RFs and averaging them across earthquakes: this will enhance the structural term while suppressing the nuisance term, which has zero conditional mean.

\section{Synthetic validation of coherent crustal phases}
\label{sec:coherency}
\setcounter{equation}{0}
\renewcommand{\theequation}{\thesection.\arabic{equation}}

In this appendix, we validate the key assumption that crustal effects are coherent within narrow bins of backazimuth and epicentral distance. This coherency justifies our conditioning strategy, where the diffusion model learns that RFs with similar condition vectors $\mathbf{c}$ share deterministic crustal responses while exhibiting variable nuisance effects.

At the crust-mantle boundary, an incident P wave converts to S waves (PS phase). Additional phases arise from reflections: PsS (reflected by free surface) and PpS (P reflected, then converted). Using the PyRaysum package~\citep{Bloch_Audet_2023}, we synthesized band-limited $\grz$ for a crustal model atop a high-velocity mantle half-space to demonstrate that crustal phase kinematics and amplitudes vary smoothly with backazimuth and epicentral distance.

\begin{figure}
\centering
\includegraphics[width=0.9\textwidth]{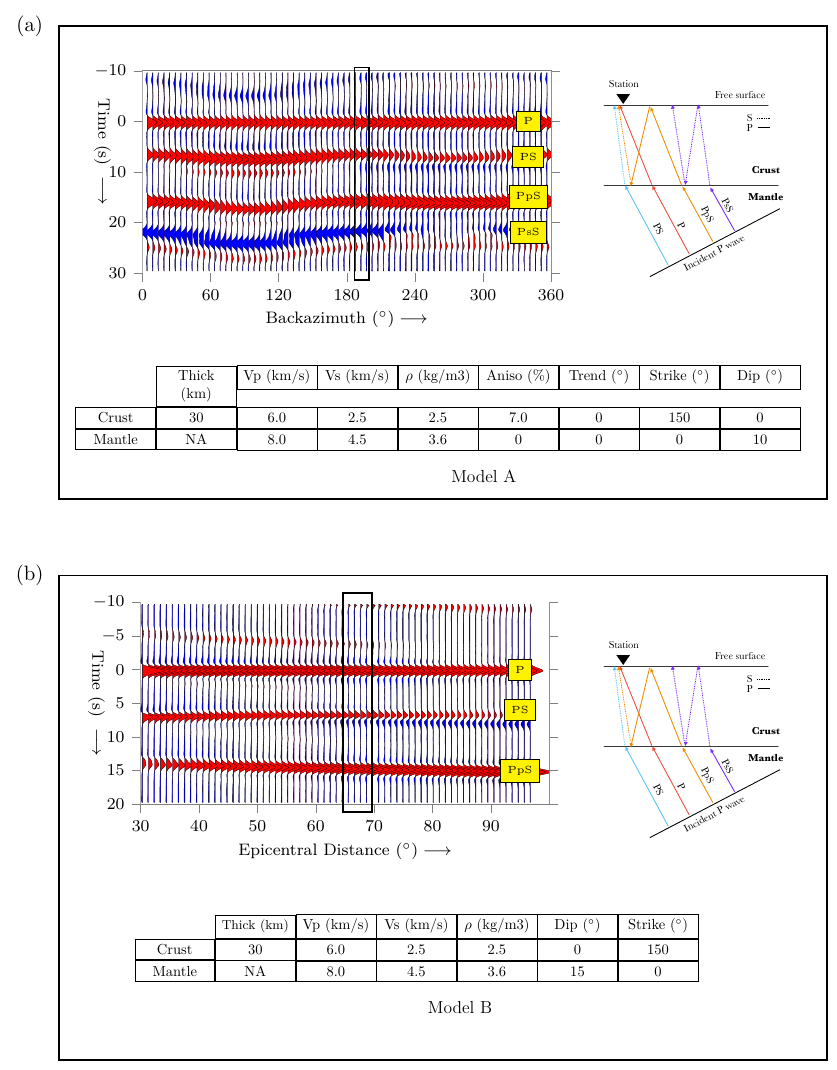}
\caption{Synthetic radial receiver functions demonstrating coherent crustal phase behavior. (a) Variation with backazimuth for anisotropic model: phases shift smoothly but remain coherent within the $10^{\circ}$ bin (rectangle). (b) Variation with epicentral distance for dipping layer: PS and PpS phases change coherently within the epicentral bin (rectangle). Raypaths (left) show how incident P waves convert to S waves at the Moho.}
\label{fig:firstplot}
\end{figure}

\textbf{Backazimuth variation:} We model an anisotropic layer ($7\%$ azimuthal anisotropy dipping $10^{\circ}$). Fig.~\ref{fig:firstplot}(a) shows radial RFs exhibit coherent phase behavior within a $10^{\circ}$ backazimuth bin (rectangle), even though phase arrival times and amplitudes change smoothly across the full range.

\textbf{Epicentral distance variation:} Fig.~\ref{fig:firstplot}(b) shows a dipping layer ($15^{\circ}$, oriented N$30^{\circ}$W). RFs change coherently within a narrow epicentral distance bin (rectangle), confirming that crustal structure impacts are minimal within our binning scheme.



\section{Conditional diffusion transformer: a deep generative model}
\label{sec:diffusion_details}
\setcounter{equation}{0}
\renewcommand{\theequation}{\thesection.\arabic{equation}}

In this appendix, we provide a detailed mathematical description of the conditional diffusion transformer used to model the conditional distribution $p_\theta(\mathbf{r} \mid \mathbf{c})$ of receiver functions.

\subsection{Forward diffusion process}

The forward diffusion process progressively adds Gaussian noise to a clean receiver function $\mathbf{r}_0$ over $T_{\max}$ timesteps, producing a sequence $\mathbf{r}_T$ for $T \in [0, T_{\max}]$. At each step, the noising process is defined as:
\begin{equation}
    \mathbf{r}_T = \sqrt{\bar{\alpha}_T} \, \mathbf{r}_0 + \sqrt{1 - \bar{\alpha}_T} \, \epsilon_T,
\end{equation}
where $\epsilon_T \sim \mathcal{N}(0, I)$ is Gaussian noise, $\alpha_s = 1 - \beta_s$ is the per-step signal retention factor, and $\bar{\alpha}_T = \prod_{s=1}^{T} \alpha_s$ is the cumulative retention factor. The variances $\beta_s$ are fixed hyperparameters that control the noise schedule. We employ a cosine-based noise schedule~\citet{nichol2021improveddenoisingdiffusionprobabilistic}. This schedule prevents excessive noise at early timesteps and maintains stable learning dynamics.

\subsection{Timestep and condition embedding}

To incorporate temporal information and conditioning variables into the neural network, we use sinusoidal positional embeddings~\citep{vaswani2023attentionneed}. For timestep $T$ and embedding dimension $d_{\text{emb}}$, the embedding is:
\begin{equation}
    \text{PE}(T, 2i) = \sin\left(\frac{T}{10000^{2i/d_{\text{emb}}}}\right), \quad \text{PE}(T, 2i+1) = \cos\left(\frac{T}{10000^{2i/d_{\text{emb}}}}\right).
\end{equation}
Similarly, condition vectors $\mathbf{c} = (\Delta, \phi, \mathbf{x}_r)$ are embedded to fixed-dimensional representations. These embeddings are added (via vector addition) to the timestep embedding and processed through dense layers to condition the denoising network.

\subsection{Transformer backbone architecture}

The denoising network $\epsilon_\theta(\mathbf{r}_T, T, \mathbf{c})$ is implemented as a Diffusion Transformer (DiT)~\citep{peebles2023scalablediffusionmodelstransformers}, which leverages the scalability and effectiveness of transformer architectures for generative modeling. In our implementation, the architecture consists of $L=4$ transformer blocks that operate on token embeddings of the input RF. The noisy RF $\mathbf{r}_T \in \mathbb{R}^{N_t}$ (with $N_t = 300$ time samples) is first converted to a sequence of tokens through patchification with patch size $p=1$, resulting in $N_t/p = 300$ tokens. Each token is then embedded into a hidden dimension $d_{\text{hidden}} = 256$ via a linear projection layer. The condition vector $\mathbf{c} = (\Delta, \phi, \text{lat}, \text{lon})$ (encoding backazimuth, epicentral distance, receiver latitude, and receiver longitude; dimensionality 4) is embedded separately and added to the timestep embedding via vector addition to modulate the transformer blocks. Each transformer block comprises multi-head self-attention layers with $H=4$ attention heads followed by feedforward networks, which process the conditioned token sequence to predict the noise $\epsilon_T$. This architecture is particularly effective for conditional generation tasks because: (i) self-attention naturally captures global dependencies in the RF across all time samples, (ii) the conditioning mechanism seamlessly integrates external information without requiring architectural modifications, and (iii) the learned representations are expressive enough to capture the complex relationship between the noisy RF and the underlying crustal structure. The use of transformers enables the model to learn rich, task-aware representations that effectively separate nuisance variability (high-variance directions) from crustal effects (low-variance directions).

\subsection{Reverse diffusion process}

The reverse process learns to denoise by predicting the noise at each timestep. Conditioned on $\mathbf{c}$, the network $\epsilon_\theta(\mathbf{r}_T, T, \mathbf{c})$ is trained to predict the noise $\epsilon_T$ in the noisy RF $\mathbf{r}_T$. At inference (sampling), the reverse process iteratively denoises from $T = T_{\max}$ down to $T = 0$:

\subsubsection{Posterior mean estimation}

Given predicted noise $\hat{\epsilon}_\theta$, we estimate the posterior mean for $\mathbf{r}_{T-1}$ using:
\begin{equation}
    \tilde{\boldsymbol{\mu}}_T = \frac{1}{\sqrt{\alpha_T}}\left(\mathbf{r}_T - \frac{\beta_T}{\sqrt{1 - \bar{\alpha}_T}} \, \hat{\epsilon}_\theta\right),
\end{equation}
where $\beta_T = 1 - \alpha_T$ is the variance at step $T$.

\subsubsection{Posterior variance}

The variance for the reverse step is computed from the analytical posterior:
\begin{equation}
    \sigma_T^2 = \frac{(1 - \bar{\alpha}_{T-1})\beta_T}{1 - \bar{\alpha}_T},
\end{equation}
which represents the uncertainty in the denoised estimate.

\subsubsection{Reverse sampling}

At each reverse timestep $T > 1$, we sample:
\begin{equation}
    \mathbf{r}_{T-1} = \tilde{\boldsymbol{\mu}}_T + \sigma_T \, \mathbf{z}_T, \quad \mathbf{z}_T \sim \mathcal{N}(0, I),
\end{equation}
where $\mathbf{z}_T$ is standard Gaussian noise. At the final step ($T = 1$), we set $\mathbf{z}_0 = 0$ to avoid adding noise.

\subsection{Training loss}

During training, for each training pair $(\mathbf{r}^{(i)}, \mathbf{c}^{(i)})$, we:
\begin{enumerate}
    \item Sample a random timestep $T$ uniformly from $\{1, \ldots, T_{\max}\}$,
    \item Sample noise $\epsilon_T^{(i)} \sim \mathcal{N}(0, I)$,
    \item Compute the noisy RF: $\mathbf{r}_T^{(i)} = \sqrt{\bar{\alpha}_T} \, \mathbf{r}_0^{(i)} + \sqrt{1 - \bar{\alpha}_T} \, \epsilon_T^{(i)}$,
    \item Compute the prediction: $\hat{\epsilon}_\theta\left(\mathbf{r}_T^{(i)}, T, \mathbf{c}^{(i)}\right)$,
    \item Minimize the mean squared error:
\end{enumerate}
\begin{equation}
    \mathcal{L} =
    \frac{1}{N} \sum_{i=1}^{N}
    \left\|
    \epsilon^{(i)}_T - \hat{\epsilon}_\theta(\mathbf{r}_T^{(i)}, T, \mathbf{c}^{(i)})
    \right\|^2.
\end{equation}

\subsection{Conditional sampling for RF generation}

After training, to generate virtual RFs for a specific condition $\mathbf{c}$, we:
\begin{enumerate}
    \item Initialize $\mathbf{v}_{T_{\max}} \sim \mathcal{N}(0, I)$,
    \item For $T = T_{\max}, \ldots, 1$:
    \begin{enumerate}
        \item Predict noise: $\hat{\epsilon}_\theta(\mathbf{v}_T, T, \mathbf{c})$,
        \item Compute posterior mean: $\tilde{\boldsymbol{\mu}}_T$ using the equation above,
        \item If $T > 1$, sample: $\mathbf{v}_{T-1} = \tilde{\boldsymbol{\mu}}_T + \sigma_T \mathbf{z}_T$,
        \item If $T = 1$, set: $\mathbf{v}_0 = \tilde{\boldsymbol{\mu}}_T$ (no noise added).
    \end{enumerate}
    \item The final output $\mathbf{v}_0$ is the generated virtual RF.
\end{enumerate}

To obtain a virtual RF with suppressed nuisance while preserving crustal effects, we generate multiple samples $\{\mathbf{v}_0^{(k)}\}_{k=1}^K$ for the same condition $\mathbf{c}$ and average them:
\begin{equation}
    \hat{\mathbf{v}}(\mathbf{c}) = \frac{1}{K} \sum_{k=1}^{K} \mathbf{v}_0^{(k)}.
\end{equation}

\section{Normalized correlation coefficient}
\label{sec:pcc}
\setcounter{equation}{0}
\renewcommand{\theequation}{\thesection.\arabic{equation}}

In this appendix, we demonstrate the computation of the normalized correlation coefficient (NCC), which is used to evaluate the quality of diffusion-generated and averaged RF.
\label{sec:syn-eval}
In the synthetic experiments, the quality of nuisance-minimized RF is determined by comparing it with the true RF. The similarity between the two is quantified by calculating the normalized correlation coefficient (NCC) between the nuisance-minimized RF, $\drec$, and the true RF, $\rrec$, at each data point $\rec_j$ as follows
\begin{equation}
    \ncc(\drec,\rrec)= \frac{{(\drec)}^\mathrm{T} \rrec}{||\drec||_{2}\,\, ||\rrec||_2},
\end{equation}
where $||\drec ||_2$ and $||\rrec ||_2$ denotes the $\ell_2$ norm of  $\drec$ and $\rrec$ respectively.

\section{Estimating crustal azimuthal anisotropy from RFs}
\label{sec:anisotropy_method}
\setcounter{equation}{0}
\renewcommand{\theequation}{\thesection.\arabic{equation}}

In this section, we describe the method used to estimate crustal azimuthal anisotropy beneath a seismic station from that station’s RFs. We model the travel‐time variations of Moho-converted S waves as a function of backazimuth using a cosine function and determine the best-fitting cosine that explains the observed backazimuthal variations in the RFs.

We assume that the crust, which exhibits azimuthal anisotropy, is flat and that the S-wave travel time for the isotropic crustal layer is denoted by $t_0$. Under these assumptions, the arrival time of the S-wave converted at the Moho can be approximated as ~\citep{liu2015crustal,Zheng2018},
\begin{equation}
t_{\mathrm{PS}}(\phi) = t_{0}- \frac{\delta t}{2}\cos(2(\psi-\phi)),
\label{eqn:cosine}
\end{equation}
where $\phi$ is the backazimuth angle of the teleseismic earthquake, $\psi$ is the azimuth of the fast-axis direction, and $t_0$ is the travel time of the S-wave for an isotropic layer. In regions with strong interface dip, multiple converting interfaces, or significant lateral heterogeneity, this simplified cosine model may produce apparent anisotropy and can bias the inferred values of $\psi$ and $\delta t$. This systematic harmonic variation allows us to estimate both the fast-axis direction and the magnitude of anisotropy beneath seismic stations using RFs. We do this by finding the best-fitting cosine function (Eq.~\ref{eqn:cosine}) to the S-wave travel times in RFs at various backazimuths. For a given station, we compute the predicted S-wave travel times at the backazimuths of RFs for the trial values of $\psi$, $\delta t$, and $t_0$ using Eq.~\ref{eqn:cosine}. We then sample the RF amplitudes at these predicted times and sum the amplitudes. This summed amplitude serves as the fitness (or goodness-of-fit) measure between the predicted anisotropy parameters ($\psi$, $\delta t$, and $t_0$) and the observed anisotropy beneath the station. Estimating the anisotropy of the RFs therefore reduces to a grid search on $\fastbz$, $\delta t$, and $t_0$ to identify the combination of parameters that yields the best fit.

For each station, we generated virtual RFs across backazimuths at an interval of $4^{\circ}$ and across epicentral distances from $35^{\circ}$ to $95^{\circ}$ at an interval of $5^{\circ}$. We then averaged the virtual RFs over epicentral distance to obtain a single virtual RF for each backazimuth. To compute linearly averaged RFs, we averaged the observed RFs over epicentral distances within each backazimuth bin. The resulting sets of virtual and linearly averaged RFs were then used to determine the values of $\fastbz$ and $\delta t$ as discussed above. The fast-axis azimuth $\fastbz$ ranges from $-90^{\circ}$ to $90^{\circ}$ in $1^{\circ}$ steps, and $\delta t$ ranges from 0 to 1.5 s in increments of 0.05 s. The isotropic travel time $t_{0}$ ranges from 2 to 7 s at an interval of 0.1 s.

Anisotropy measurement techniques require high‑quality receiver functions with broad back‑azimuth coverage to ensure robust results. Contamination by nuisance effects—such as instrumental noise, source-receiver path effects, and scattering—can introduce spurious arrival-time picks and bias anisotropy estimates. By generating virtual receiver functions with suppressed nuisance variability, the conditional diffusion transformer approach enhances the reliability and precision of the crustal anisotropy measurements, particularly in regions with limited earthquakes.

\end{document}


\maketitle
\begin{center}
\underline{\textbf{\huge Supplementary material}}    
\end{center}

\maketitle
\section{Introduction}
This Supplementary Information provides additional methodological details, extended analyzes, and supporting data that complement the main manuscript. The figures present expanded visualizations of key results, including detailed comparisons and illustrative examples that clarify and substantiate the main findings. It is composed of the following sections:
\begin{itemize}
    \item  Examples of RF samples from the diffusion model. 
    \item  RFs at Cascadia station SNB from our method and  \textit{Bloch et al., 2023}.
    \item  Estimation of azimuthal anisotropy at station DSC in southern California.
    \item  Transverse RFs from our method and linear averaging for stations transecting major faults in southern California.
\end{itemize}


\section{Sampled RFs from conditional diffusion model}

\begin{figure}
\centering
\includegraphics[width=\textwidth]{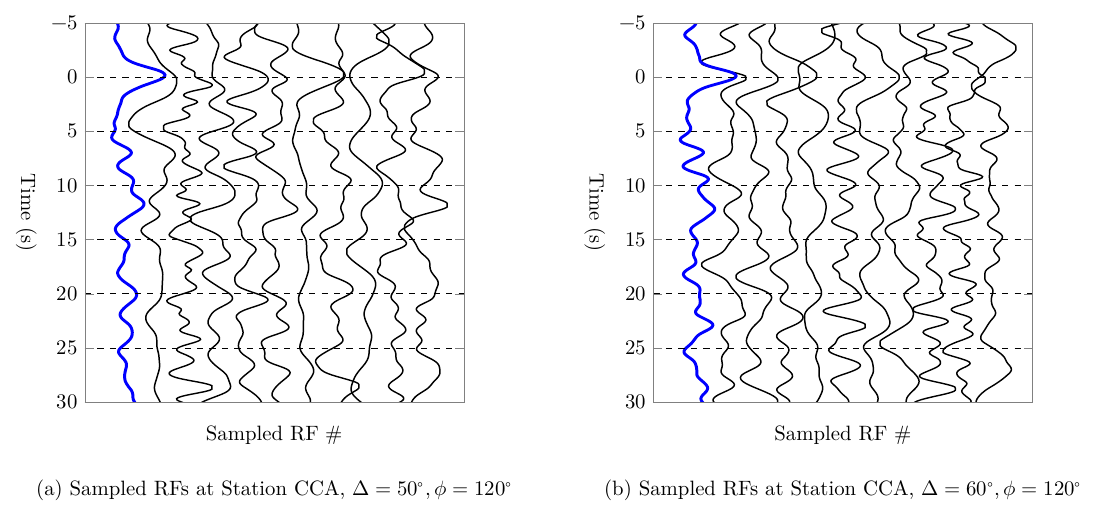}
\caption{This figure shows the sampled RFs from two conditions: (a) Station CCA, $\Delta=50^{\circ},\phi=120^{\circ}$ and (b) Station CCA, $\Delta=60^{\circ},\phi=120^{\circ}$. The blue line represents the averaged RF for each condition. Note the similarity of the two averaged RFs, which are at closeby epicentral distance and variation of nuisance effects across the sampled RFs. Only 10 RF samples are shown here.}
\label{fig:sampleRFs}
\end{figure}

After training, we can sample many realizations of RFs for a given condition vector (i.e., back-azimuth angle, epicentral distance, and station location coordinates). These RFs, corresponding to the same condition vector, exhibit varying nuisance effects. However, when we average all such RFs, the nuisance effects are suppressed. The resulting averaged RF represents a virtual RF that predominantly reflects crustal effects. Figure~\ref{fig:sampleRFs} illustrates this process at a real seismic station.

\section{RFs from our method and previous study}

\begin{figure}

\includegraphics[trim={0cm 0 0.0cm 0cm},clip,width=\textwidth]{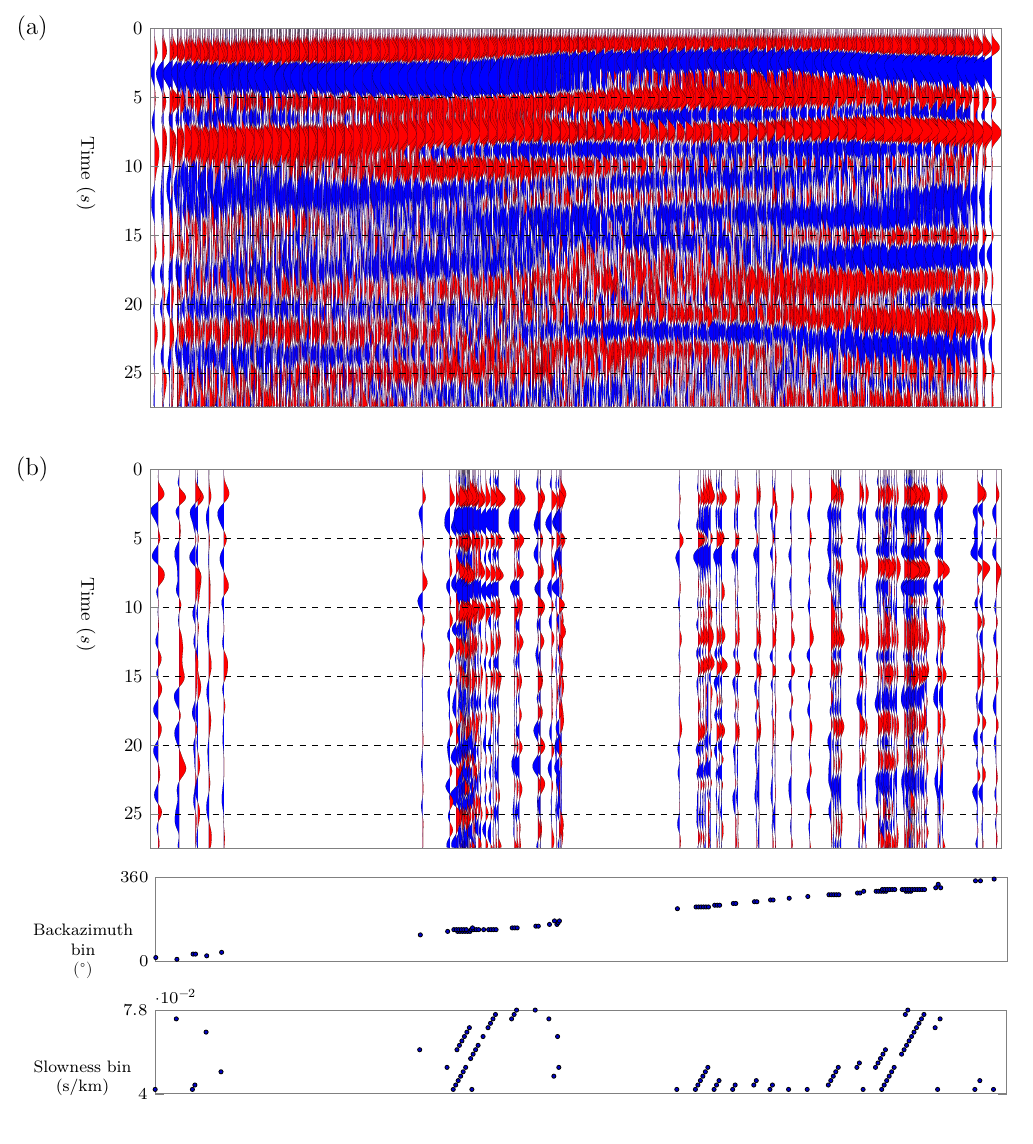}

\caption{Radial RFs at Cascadia station SNB from (a) our method and (b) \textit{Bloch et al. [2023]} demonstrating the full coverage of RFs across backazimuth angle and slowness. The RFs are organized by ascending backazimuth and slowness. Converted S-waves between 0 to 10s time window are more prominent in (a) compared to (b).}
\label{fig:snbrad}
\end{figure}

\begin{figure}

\includegraphics[trim={0cm 0 0.0cm 0cm},clip,width=\textwidth]{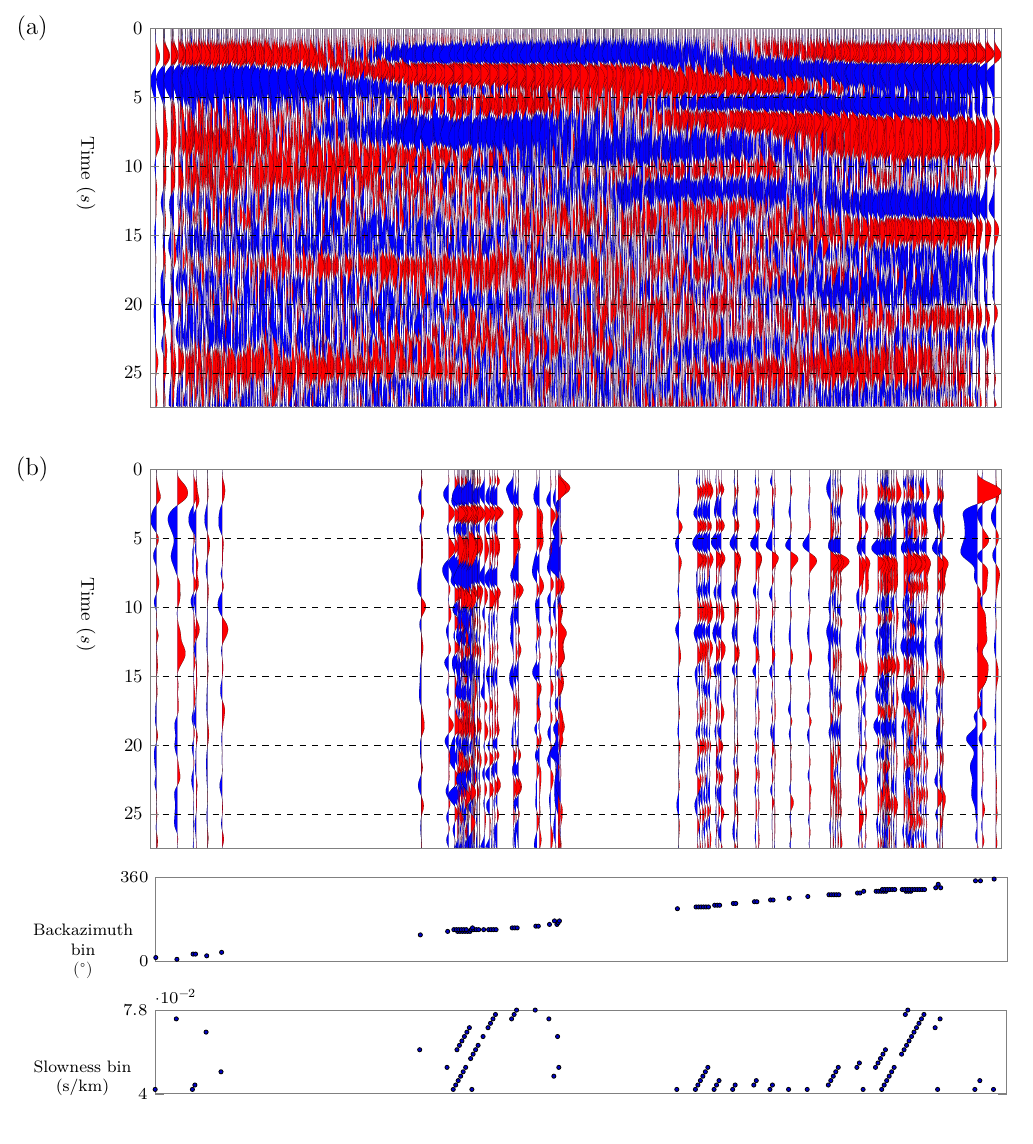}

\caption{Similar to Figure~\ref{fig:snbrad} but for transverse RFs. Note the polarity reversal is clearly observed between 0 to 5 s in (a) virtual RFs, approximately at backazimuth $60^{\circ}$ and $180^{\circ}$}
\label{fig:snbtran}
\end{figure}

Traditional RF analyses focus on selecting high‑quality RFs using various criteria and then averaging them within specific back‑azimuth and epicentral‑distance bins. In contrast, our method incorporates information from all RFs, regardless of signal quality, and generates virtual RFs that span the full range of back‑azimuths and epicentral distances. Figures~\ref{fig:snbrad} and~\ref{fig:snbtran} show the radial and transverse RFs at Cascadia station SNB produced by our method and by \textit{Bloch et al.} (2023).

\section{Transverse RFs for station profile in southern California}

\begin{figure}
\centering
\includegraphics[width=\textwidth]{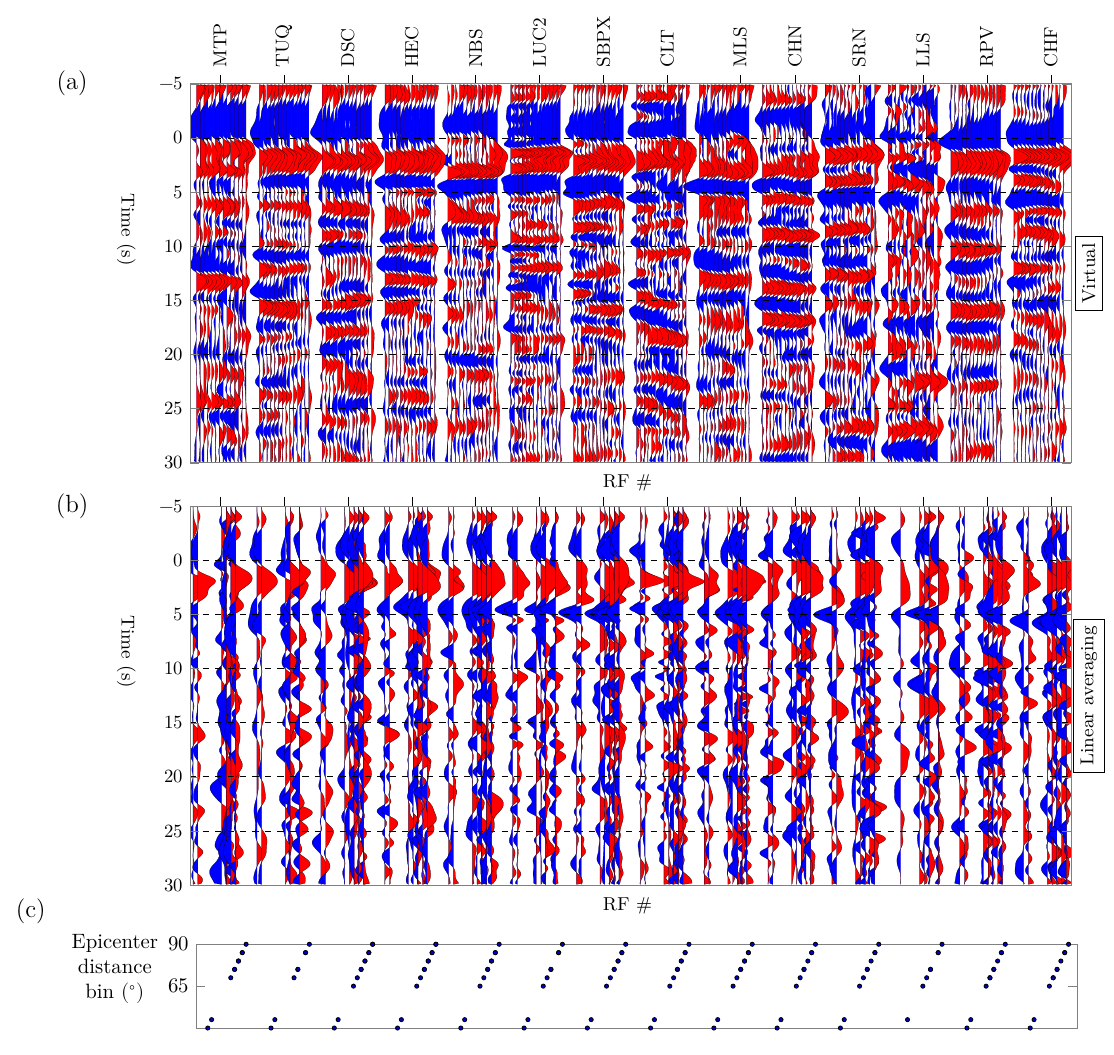}
\caption{(a) Virtual and (b) linearly averaged transverse RFs for a station profile transecting the San Andreas Fault (SAF), San Jacinto Fault Zone (SJFZ) and Elsinore Fault (EF) in Southern California. Station locations are given in Fig.~\ref{fig:calimap}. At each station, RFs from different epicentral distance are plotted. Backazimuth angle is fixed at $304^{\circ}$. (c) shows epicentral distance bin corresponding to (b) linearly averaged RFs. Virtual RFs are similar to linearly averaged RFs but show higher clarity across all bins and stations.}
\label{fig:calitran}
\end{figure}

\begin{figure}
\centering
\includegraphics[width=\textwidth]{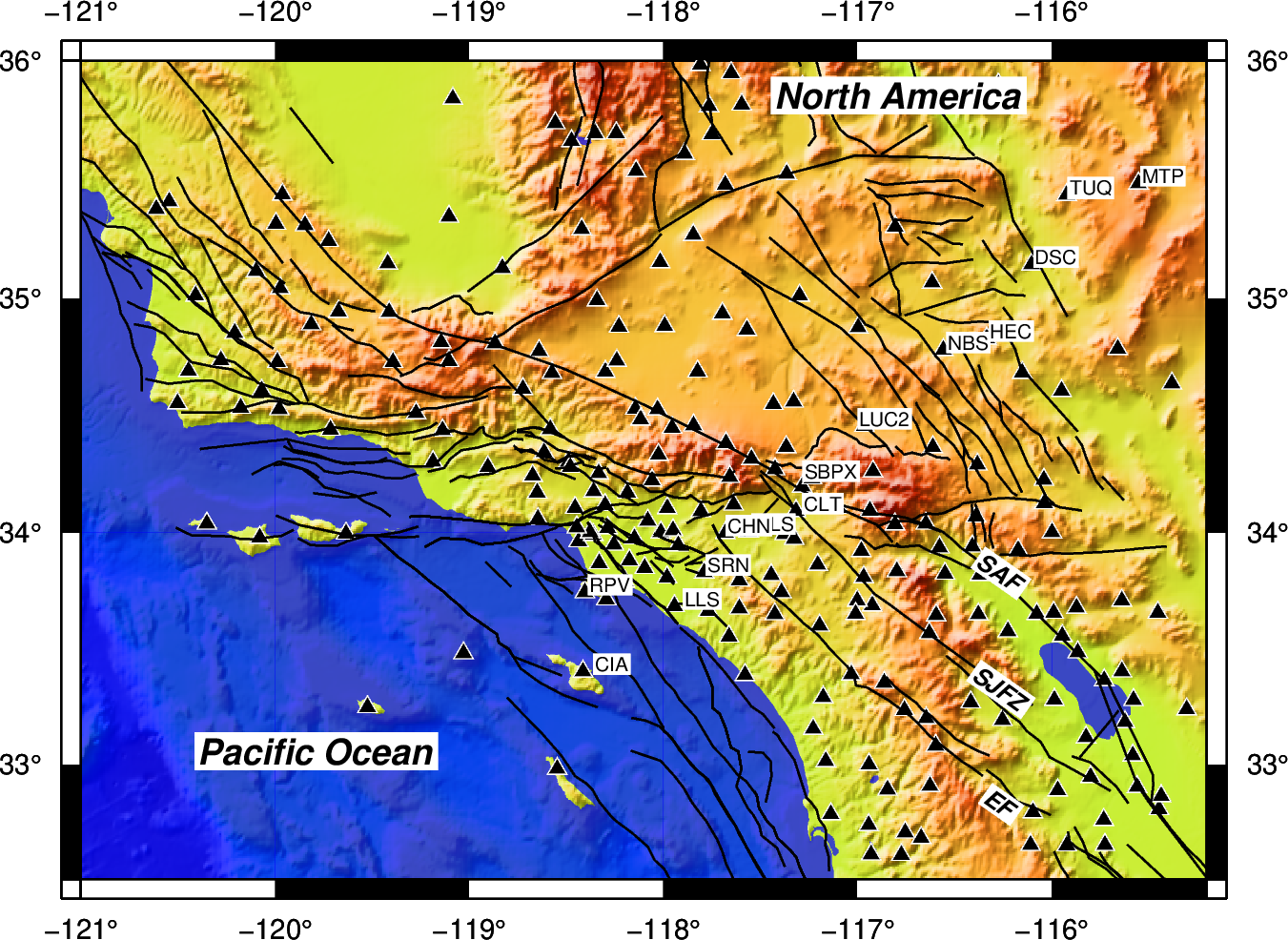}
\caption{Map showing the seismic station profile transecting the San Andreas Fault (SAF). San Jacinto Fault (SJF) and Elsinore Fault (EF) are also indicated}
\label{fig:calimap}
\end{figure}

High-quality RFs are challenging to derive from transverse components because the seismic signal is weak and largely masked by noise and other nuisance effects. However, they contain critical information about crustal structure, especially regarding dip and anisotropy. Our method yields a marked improvement in transverse RFs relative to simple linear stacking. Fig.~\ref{fig:calitran} presents transverse RFs obtained with our approach and with linear averaging for a station profile that intersects major faults in southern California. The location of these seismic stations is given in Fig.~\ref{fig:calimap}

\section{Estimation of anisotropy at station DSC in southern California}

\begin{figure}
\centering
\includegraphics[width=\textwidth]{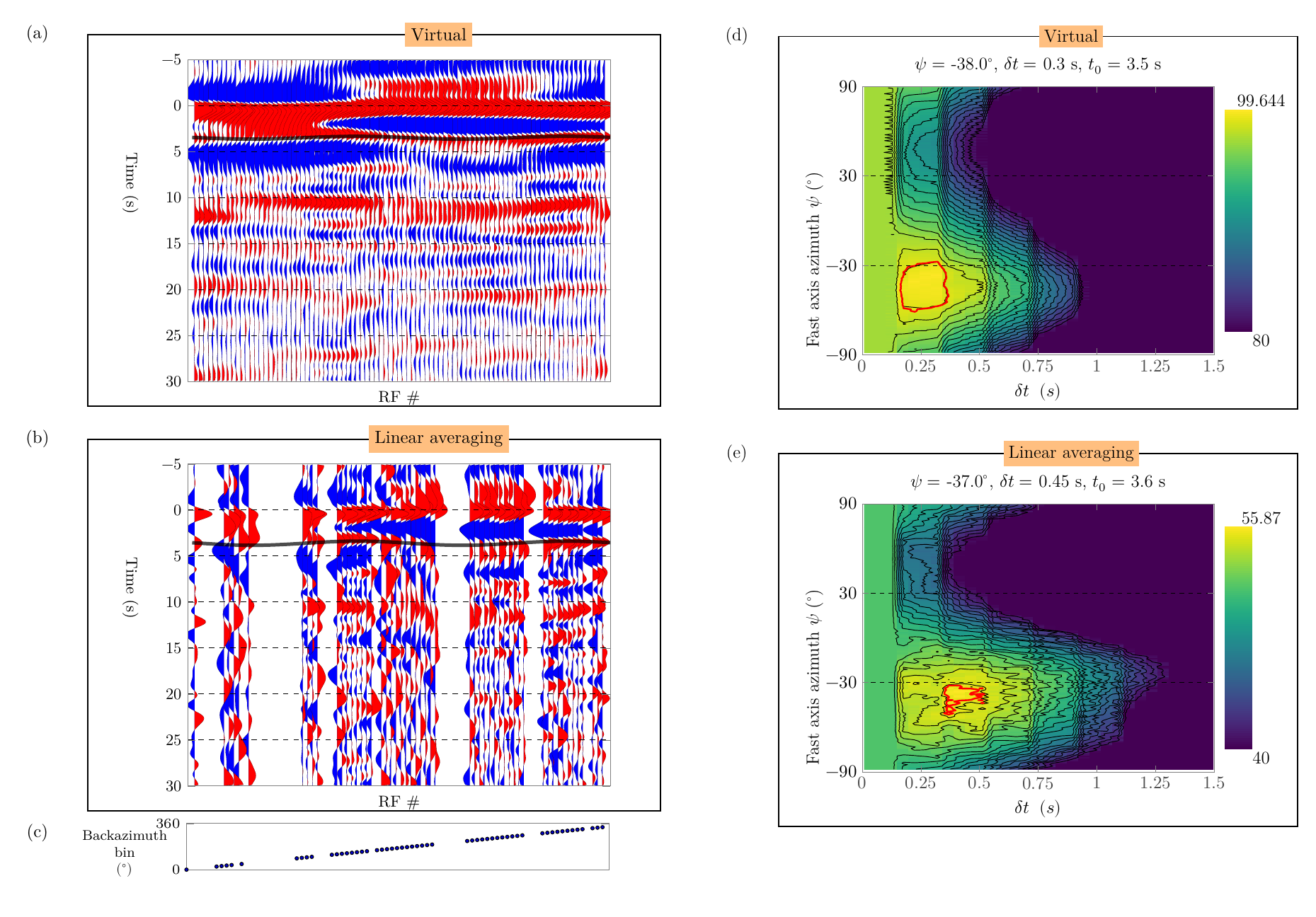}
\caption{This figure shows (a) virtual RFs generated from the diffusion transformer and (b) linearly averaged RFs at seismic station DSC in southern California (location is given in Fig.~\ref{fig:calimap}). Panel (c) shows the backazimuth bins corresponding to the RFs in (b). Panels (d) and (e) show the estimated anisotropy parameters ($\psi$,$\delta t$ and $t_{0}$) using (a) and (b), respectively. The travel time of converted S-wave corresponding to the estimated parameters is indicated by solid black line in (a) and (b). The full backazimuth coverage of the virtual RFs leads to improved parameter estimation, as indicated by the evenly spaced contours. The red lines in (d) and (e) denote the 98\% confidence regions.}
\label{fig:dsc}
\end{figure}

We estimated azimuthal anisotropy at seismic stations in southern California using the method described in the main text. Anisotropy estimates based on virtual radial receiver functions (RFs) are better constrained and more reliable because nuisance effects are reduced and backazimuthal coverage is improved. This is corroborated by the distribution of the misfit for different trial values of backazimuth angle ($\psi$) and anisotropy amplitude ($\delta t$) at selected isotropic travel times ($t_0$). To further support the results presented in the main text, we show the corresponding plot for station DSC in Fig.~\ref{fig:dsc}.

